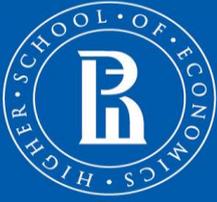

*Fuad Aleskerov, Natalia Meshcheryakova,*
*Alisa Nikitina, Sergey Shvydun*

# KEY BORROWERS DETECTION BY LONG-RANGE INTERACTIONS

BASIC RESEARCH PROGRAM

WORKING PAPERS

SERIES: FINANCIAL ECONOMICS
WP BRP 56/FE/2016

*Fuad Aleskerov[a], Natalia Meshcheryakova[b],*

*Alisa Nikitina[c], Sergey Shvydun[d]*

# KEY BORROWERS DETECTION
# BY LONG-RANGE INTERACTIONS[e]

We propose a new method for assessing agents' influence in financial network structures, which takes into consideration the intensity of interactions. A distinctive feature of this approach is that it considers not only direct interactions of agents of the first level and indirect interactions of the second level, but also long-range indirect interactions. At the same time we take into account the attributes of agents as well as the possibility of impact to a single agent from a group of other agents. This approach helps us to identify systemically important elements which cannot be detected by classical centrality measures or other indices. The proposed method was used to analyze the banking foreign claims for the end of 1Q 2015. Under the approach, two types of key borrowers were detected: a) major players with high ratings and positive credit history; b) intermediary players, which have a great scale of financial activities through the organization of favorable investment conditions and positive business climate.

JEL Classification: C7, G2.

Keywords: systemic risk, key borrower, interconnectedness, centrality.

[a]National Research University Higher School of Economics (HSE), International Laboratory of Decision Choice and Analysis, V.A. Trapeznikov Institute of Control Sciences of Russian Academy of Sciences (ICS RAS), Moscow, alesk@hse.ru

[b]National Research University Higher School of Economics (HSE), International Laboratory of Decision Choice and Analysis, V.A. Trapeznikov Institute of Control Sciences of Russian Academy of Sciences (ICS RAS), Moscow, natamesc@gmail.com

[c]National Research University Higher School of Economics (HSE), International Laboratory of Decision Choice and Analysis, Moscow, knurova.a.a@gmail.com

[d]National Research University Higher School of Economics (HSE), International Laboratory of Decision Choice and Analysis, V.A. Trapeznikov Institute of Control Sciences of Russian Academy of Sciences (ICS RAS), Moscow, shvydun@hse.ru

[e]The paper was prepared within the framework of the Basic Research Program at the National Research University Higher School of Economics (HSE) and supported within the framework of a subsidy by the Russian Academic Excellence Project '5-100'. The work was conducted by the International Laboratory of Decision Choice and Analysis (DeCAn Lab) of the National Research University Higher School of Economics and the Laboratory №25 of V.A. Trapeznikov Institute of Control Sciences of Russian Academy of Sciences

# Introduction

The financial crisis has reminded how important it is to look at the links and connections inside the financial system. We saw that major disruptions such as failure or a near failure of certain financial institutions rapidly spilled over to the whole system. In other words, there is a need to detect systemically important financial institutions or countries that require more careful investigations and observations. This issue has significant implications for macro-prudential surveillance, and hence for financial stability in terms of systemic risk identification.

Pivotal agent identification recently has received particular attention in the context of the risk allocation problem. Existing studies use different approaches to identify key players in financial markets. Most of them are based on quantitative statistical analysis of indicators for each element or on the analysis of the sustainability of networks.

The most traditional approach related to systemic importance assessment is an indicator-based approach. It is usually based on the system of financial indicators, estimated by a regulator and relied on measuring the banks involvement in certain activities. Another common approach in the literature, which is focused on the evaluation of the contribution of financial institutions to cumulative systemic risk, is called a structural approach. It is based on the micro-founded, general equilibrium theoretical framework, which indicates that financial instability can arise either through systemic shocks, contagion after idiosyncratic shocks or through a combination of both.

Another approach of estimating the degree of systemic risk is to apply a network theory. In that case a network can be represented as the system of nodes (financial institutions) and links (flows of capital) among them. For the purpose of measuring the degree of importance in networks many centrality indices were proposed.

We use the network analysis, and try to take into consideration all the drawbacks of existing indices and centrality measures and to design a new method for assessing the significance of elements of financial systems based on the intensities of long-range interactions. In fact, our methodology is an attempt to improve the approach proposed in (Aleskerov et al., 2014), which does not take into account long-range interactions between elements of the network.

The paper is organized as follows. In Section 1, we present a detailed review of the literature related to different approaches of systemic risk assessment. In Section 2we emphasize the drawbacks of existing approaches and provide a small example to describe the basic idea of our methodology. In Section 3, we formally describe the proposed model and present more complicated example. Finally, we demonstrate the empirical application of our model to the data of international lending activities of banks' head offices of different countries.

# 1. Literature review

In this Section let us briefly review quantitative models of systemic risk assessment which have been proposed in the literature.

In most contemporary studies, systemic risk is understood in two different but related ways. First, the "systemic risk contribution" is associated with a large and complex financial institutions which correspond to a negative externality of the risk for other economic agents. On the other hand, systemic risk is often understood as financial system risk. We follow the second idea which is the analogue of the assessment of the total risk measure, its composition and the development of systemic risk assessment for monitoring purposes.

Systemic risk can be measured for markets as a whole or at the financial institutions level to identify systemically important elements. For both markets and institutions, these measures may be direct, using analytical models, or indirect, using indicators that are considered to relate to systemic risk.

An example of the indirect measure for markets is an *indicator-based approach*, which usually used in cases when we do not have an exogenous measure of systemic risk and have to rely on expert judgments and priors to validate the model-based measures. For example, the (ECB, 2015), (IMF, 2015), (BCBS, 2013) proposed indicators for size, interconnectedness, and substitutability to measure the systemic importance of a financial institution. In (Thomson, 2009) the use of size and the four C's (contagion, concentration, correlation, and conditions) as criteria to determine the systemic importance was p. Similarly, (Patro et al., 2013) proposed that stock return correlation is a useful indicator of systemic risk for markets as a whole: while holding the firm level probability of default constant, a higher correlation implies a higher joint probability of default for the system, such that correlation can serve as a useful indicator of systemic risk.

There is also a considerable amount of literature that has investigated more complex models for systemic risk, so called *structural approach*. It develops the ideas of (Gray et al., 2008) who propose the use of contingent claims analysis to evaluate the sensitivity of an economic system's balance sheets to external shock. Similarly, a direct model-based measure of the systemic importance was presented in (Segoviano and Goodhart, 2009), in which joint probability of distress as well as a banking stability index for the financial sector was estimated and in (Adrian, Brunnermeier, 2009) with CoVaR measure. Other examples of works under this approach include (Allenspach and Monnin, 2006), (Goodhart et al. 2006), (Hartmann et al., 2005), (Huang et al., 2009), (Zhou, 2010), (Chan-Lau, 2010).

As we have mentioned in previous Sections, there is a third strand of the literature investigating systemic risk allocation. There is a broad range of works employing the network theory. *Network analysis* of economic and financial structures has already been applied in the context of stock ownership networks (Garlaschelli et al., 2005), emergence of contagion and systemic risk in the interbank market and in payment systems (Angelini et al., 1996; Furfine, 2003; Iori et al., 2006) and also in terms of how interconnected a financial system can be at the national and international level (Allen, Babus, 2009; Allen, Gale, 2000). We extend this part of literature with our current study.

For the purposes of measuring the degree of importance in networks many *centrality indices* were proposed (Bonacich, 1972), (Barrat et al., 2004), (von Peter, 2007), (Newman, 2010). The three most widely used centrality measures are degree, closeness, and betweenness (Freeman 1977, 1979). Some measures differ in their use of undirected and directed graphs.

*Degree centrality* refers to the number of ties a node has to other nodes. Nodes have higher centrality to the extent it can gain access to and/or influence over others. A central node occupies a structural position (network location) that serves as a source or a channel for larger volumes of information exchange or other resource transactions with other nodes.

Degree centrality $C_i^d(g)$ of node $i$ in network $g$ is calculated as

$$C_i^d(g) = \frac{\eta_i(g)}{n-1} = \frac{|N_i(g)|}{n-1} \in [0,1],$$

where $\eta_{i(g)}$ is the degree of $i$ in adjacency matrix $g = [g_{ij}]$ and $N_i(g)$ is the set of neighbors of $i$ in $g$, n is the number of vertices.

Degree centrality for a directed graph has one of two forms: in-degree centrality and out-degree centrality. Accordingly, in-degree centrality is evaluated using the number of in-coming links to the node, and out-degree uses the number of out-going links from the node.

For systemic risk analysis and risk allocation in lending activities *weighted in-degree centrality* (WInDeg) indicates the most active borrowers on the market. It is important to emphasize that this measure lacks information on the number of links of the neighbors of each borrower. Two borrowers can take the same amount of money, but if one borrower is connected to more lenders, the contagion effect from the failure of this borrower would be higher compared to the failure of the other borrower.

*Weighted out-degree centrality* (WOutDeg) indicates the most active lender, and for risk analysis represents borrowers with the largest financial resources and attractive financial instruments to invest. As a previous measure, this one do not take into consideration information about the number of links. *Weighted degree* centrality is simply the number of other nodes connected directly to a node taking into account the weights on the edges. It is an indicator of a lender's financial activity and shows the level of its involvement in all types of financial transactions.

The value of *weighted degree difference*, which could be calculated as *Weighted out-degree centrality - weighted in-degree centrality,* shows the aggregated role of a particular node in the system. As a result for systemic risk analysis, all elements could be divided into two groups: net creditors (when *WInDeg>WOutDeg*) or net borrower (*WInDeg<WOutDeg*). Low values of this measure in case of banking foreign claims analysis can be explained by two factors: attractive conditions for direct foreign investments or the realization of government financial assistance programs. In both cases, the incoming flow will be significantly higher than outgoing flows.

*Betweenness centrality* show often the vertex is placed on the shortest path between any two nodes in the network. Betweenness centrality for node *i* is the sum of the proportions for all pairs of actors *j* and *k*, in which actor *i* is involved in a pair's geodesic(s), i.e.

$$C_i^b(g) = \sum_{j<k} \frac{\sigma_{jk}(n_i)}{\sigma_{jk}}$$

where $\sigma_{jk}(n_i)$ is the number of geodesics between k and j containing $i \notin \{k,j\}$, $\sigma_{jk}$ is the total number of geodesics between k and j.

In our work it will be constructed in a way that maximizes the total sum on the edges on the shortest path. In terms of loans it could be interpreted as a measure of how often the borrower is on the most popular capital transition channel (the path with the largest capital flow between any pairs of borrowers).

*The level of closeness* shows how close the node is located to other nodes or how easy we can reach other nodes in the network from a particular one. It could be defined as the inverse of farness, which in turn is the sum of distances to all other nodes in undirected graph

$$C_i^C(g) = \frac{1}{\sum_{i \neq j} l_{ij}(g)}$$

where $l_{ij}(g)$ is the geodesic distance between *i* and *j* in *g*.

Thus, the more central a node is the lower its total distance from all other nodes. Note that taking distances from or to all other nodes is irrelevant in undirected graphs, whereas in directed graphs distances to a node are considered a more meaningful measure of centrality.

In our work it will be constructed in a way that maximizes total capital flow on the path and shows how easy it is to reach the particular node in a network from the other nodes. In terms of loans it shows the most efficient financial intermediaries, which attract high volume of resources.

*Eigenvector centrality* is a measure of the influence of a node in a network. It assigns relative scores to all nodes in the network based on the concept that connections to high-scoring nodes contribute more to the score of the node in question than equal connections to low-scoring nodes. The value is obtained by solving the linear system $v = A'v$, where $v$ is the vector of the importance scores for $i$ and adjacency matrix $A = [a_{ij}]$. The solution is represented by the eigenvector corresponding to the eigenvalue 1.

*PageRank* is a version of the eigenvector centrality measure. There are different versions of the measure. In one version, the value of a network node centrality is calculated by the following formula

$$PR(i) = (1 - \alpha) + \alpha \cdot \sum_j \frac{PR(j \to i)}{|N_i(g)|}$$

where $PR(i)$ is the value of the node centrality $i$, $\alpha$ is so-called damping factor (usually set at 0,85), $PR(j \to i)$ is the value of the central node $j$, which has a direct link with the node $i$, $N_i(g)$ is the number of links of a node $j$. In other words, PageRank for node $i$ is determined by the PageRank of each node $j$, which has a direct link with the node $i$.

Thus, we can see that some centrality measures are based on the number of links to other nodes with/without respect to their importance (e.g. degree, eigenvector measures). Other techniques consider how close each node is located to other nodes in terms of the distance (e.g., the closeness measure) or how many times it is on the shortest paths connecting different node-pairs (e.g., the betweenness measure), etc.

Other attempts to evaluate the degree of importance of elements in networks are based on simulation mechanism. For example, a method of firm dynamics simulations was developed by applying game theory to a stochastic agent model in order to analyze the uncertainty in the business environment (Ikeda et al, 2007), (Giovanetti, 2012). Also simulation procedure is widely used to assess the industrial transactions networks, property relations, and the dynamics of industrial and innovation clusters as well as modeling the financial risks in the interbank market and payment systems.

In (Leonidov, Rumjanzev, 2013) the authors analyze the Russian interbank network structure based on such characteristics as borrowers and lenders distribution, assets and liabilities distribution and the relationship between the ant of claims with a number of counterparties. Their methodology of the analysis of systemic risk is based on a study of the consequences of default of one of the banks considered as a result of simulation of cascade defaults on interbank network similar to the process of contagion effect.

Many attempts of key nodes detection in networks came from *cooperative game theory*. In that case, the network is interpreted as a set of interacting individuals that contribute to a total productive value of the network and the problem is how to share generated value among them. In (Myerson, 1977) there was proposed a measure which is based on the Shapley-Shubik index (Shapley, Shubik, 1954) for communication games. The Myerson value is an allocation rule in the context of network games where the value of each individual depends on the value generated by the network with and without that individual. Several attempts to employ power indices to find systemically important financial institutions were accomplished (Tarashev et al, 2010), (Drehmann, Tarashev, 2011), (Garratt, 2012). However, the above-described techniques do not fully consider the intensity of connections among individuals.

In (Aleskerov, 2006) a novel approach for estimating the intensities of agents' interactions was proposed. This method is based on the power index analysis and it was used to find the most pivotal fractions in Russian Parliament (1999-2003). Later, this method has been adapted to analyze the effects of the network and used in (Aleskerov et al., 2014). In that work only short-range interactions were taken into account and we will solve this problem in next Sections.

## 2. Numerical Example 1

In this Section we demonstrate some shortages of the existing methods of systemic risk assessment in details and propose a new method to solve them.

Consider a hypothetical Numerical Example 1 of agents' lending activities (graphical representation is shown on Figure 1) with a complex system of interconnections. There are 10 agents in the system, one of them is a pure lender, six of them are both lenders and borrowers while three remaining elements are pure borrowers. In this case, the network structure is chosen such that the ratio of elements that play different roles (pure lenders, pure borrowers, both lenders and borrowers) was in line with the so-called "bow-tie picture" (Strogatz et al., 2001; Leonidov, Rumyantsev, 2013), which allows to display a general structure of a directed graph. This structure allows studying the quantitative composition of the components and the connections between them in terms of emergence of cascade defaults and contagion effects.

The values on the edges represent the amount of loan (in USD) that one agent gave to another one. Arrows in the network indicate the direction of the money flows. For instance, the agent 2 borrows $500 from the agent 1 and at the same time lends $40 to the agent 3, $100 to the agent 6 and $60 to the agent 9.

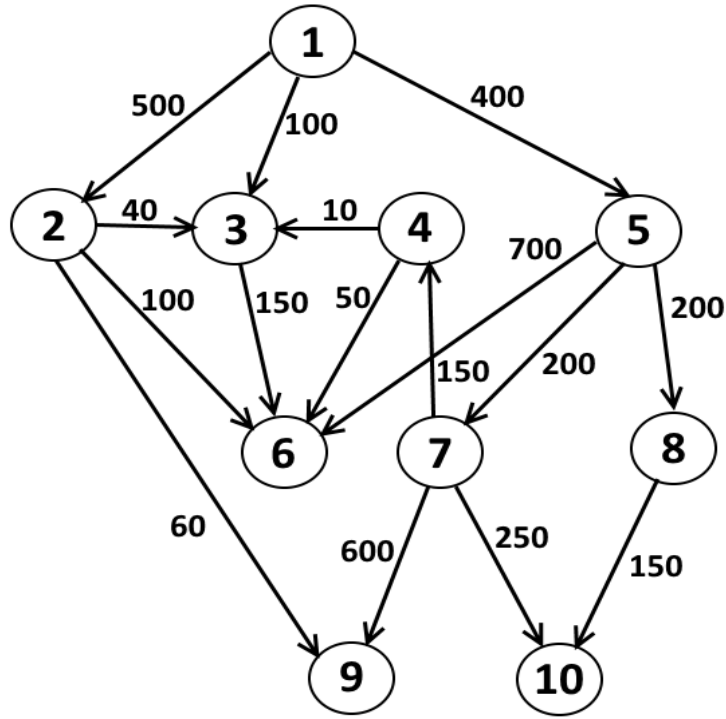

**Figure1**. Numerical Example 1

The standard approach for key elements detection in the network is to calculate different centrality measures. In Table 1 we provide the topological importance of each element by classical centrality measures that reveal the most pivotal borrower in the system. The definition of centrality measures is given in the last Section of the paper.

**Table 1.** Classical centrality measures for Numerical Example 1

| Indices\Agents | 1 | 2 | 3 | 4 | 5 | 6 | 7 | 8 | 9 | 10 |
|---|---|---|---|---|---|---|---|---|---|---|
| Weighted In-degree | 0 | **500** | 150 | 150 | 400 | **1000** | 200 | 200 | **660** | 400 |
| Weighted Out-degree | 1000 | 200 | 150 | 60 | 1100 | 0 | 1000 | 150 | 0 | 0 |
| Weighted Degree Difference | 1000 | -300 | 0 | -90 | 700 | **-1000** | **-800** | -50 | **-660** | -400 |
| Weighted Degree | **1000** | 700 | 300 | 210 | **1500** | 1000 | **1200** | 350 | 660 | 400 |
| Closeness, in | **0.0111** | **0.0036** | 0.0002 | 0.0003 | **0.0026** | 0.0003 | 0.001 | 0.001 | 0.0004 | 0.0003 |
| Closeness, out | 0,0002 | 0,001 | 0,0016 | 0,001 | 0,0002 | 0,0(1) | 0,0003 | 0,0016 | 0,0(1) | 0,0(1) |
| Betweenness | 0 | 1 | 0 | **3** | **5** | 0 | **6** | 0 | 0 | 0 |
| Eigenvector | **0.67** | 0.46 | 0.21 | 0.11 | **1.00** | **0.81** | 0.45 | 0.23 | 0.31 | 0.15 |
| PageRank | 0.06 | 0.08 | 0.09 | 0.07 | 0.08 | **0.25** | 0.07 | 0.07 | **0.11** | **0.13** |

As it is shown in Table 1, centrality measures give results, which are very dissimilar in comparison to each other. Analyzing the measures, we can conclude that classical centrality indices consider borrowers 6 and 9 as the most powerful. However, none of these measures except PageRank considers the lender 10 as the key borrower, whereas in fact, its bankruptcy can start a chain reaction that could lead to bankruptcy of the lenders 5 and 1. Look at the results from Table 2. Here we provide the consequences for individual bankruptcies of elements 1-10 for 3 levels. On each step we are moving along the chain of borrowers checking the bankruptcy condition (in this case it is equivalent to losing more than 25% of their assets) .

**Table 2.** Multistep simulation procedure for Numerical Example 1

| Element, $i$ | Total Losses/Total Credits Provided* | | | Bankrupted Elements | | |
|---|---|---|---|---|---|---|
| | Step 1 | Step 2 | Step 3 | Step 1 | Step 2 | Step 3 |
| 1 | - | - | - | ∅ | - | - |
| 2 | **500/1000** | - | - | {1} | - | - |
| 3 | 100/1000; 40/200; 10/60 | - | - | ∅ | - | - |
| 4 | 150/1000 | | | ∅ | | |
| 5 | **400/400** | | | {1} | | |
| 6 | **100/200**; **150/150**; **50/60**; **700/1100** | - | - | {2, 3, 4, 5} | {1} | - |
| 7 | 200/1100 | - | - | ∅ | - | - |
| 8 | 200/1100 | - | - | ∅ | - | - |
| 9 | 60/200; **600/1000** | 200/1100 | - | {7} | - | - |
| 10 | **150/150**; **250/1000** | **400/1100** | 400/400 | {7,8} | {5} | {1} |

\* - the value represents total losses from element $i$ bankruptcy divided by total amount of loans granted by each of $i$'s direct creditors. Number of element indicated in the "Bankrupted elements" column.

One of the reason why centrality indices do not reveal the lender 10 as pivotal is that they take into account the number of direct interactions between agents, but at the same time ignore information about indirect links. Moreover, it is not always reasonable to consider all links in the particular network. Next, we will show that some borrowers may affect the financial stability of the lender only in conjunction with others, forming a so-called critical groups.

In (Aleskerov et al., 2014) a novel method for estimating the intensities of agents' interactions was proposed. This method is based on the power index analysis that was worked out in (Aleskerov, 2006) and adjusted for the network theory. The index is called a Key Borrower Index (KBI) and is employed to find the most pivotal borrower in a loan market in order to take into account some specific characteristics of financial interactions. An important feature of the KBI  is that it uses the parameter $q$ which varies with agent and represents its critical amount of loan.

For the "one lender, many borrowers" case the KBI is calculated for each lender individually in order to determine the influence of each borrower to him. The distinct feature of the proposed index is that it takes into account short-range interactions between each lender and its borrowers. In other words, only direct neighbors are considered to estimate the direct and indirect influence to a specific lender. The intensity of direct connections $p_{Li}$ between the lender $L$ and the borrower $B_i$ is calculated as

$$p_{Li} = \frac{c_{Li}}{\sum_k c_{Lk}},$$

where $c_{Lk}$ is the amount of loan from the lender $L$ to the borrower $B_k$, while the intensity of indirect connections $p_{ji}$ between the lender $L$ and the borrower $B_i$ through $B_j$ is calculated as

$$p_{ji} = \begin{cases} \frac{c_{ji}}{\sum_k c_{Lk}}, if \; c_{Lj} > 0, c_{ji} < c_{Li} \text{ and } i \neq j, \\ \frac{c_{Li}}{\sum_k c_{Lk}}, if \; c_{Lj} > 0, c_{ji} > c_{Li} \text{ and } i \neq j, \\ 0, otherwise. \end{cases}$$

After the intensity of connections between the lender $L$ and its borrowers is calculated, a set of all possible critical groups of borrowers for the lender $L$ is constructed. A group of borrowers is critical if the total loan taken by these borrowers from the lender $L$ is more than or equal to some pre-defined threshold $q_i$. The critical group is interpreted as a group whose default may lead to the default of the lender (while the lender is able to cover its losses from the distress of members outside the critical group). Thus, the group is "critical" if the total amount of its members' borrowings is greater than or equal to a predefined threshold $q_i$.

After a set of critical groups for the lender $L$ is defined, we can identify a total number of groups where each borrower $B_i$ plays a pivotal role. A borrower $B_i$ is pivotal in the group if his/her exclusion from the critical group makes it non-critical. The value of the index for each borrower reflects the magnitude of his/her pivotal role in the group. The higher the value, the more pivotal the agent is. The most pivotal borrower will be the one that becomes pivotal in more critical groups than any other borrower does.

The total intensity of connection between the lender $L$ and the borrower $B_i$ is aggregated over the intensities of all groups where the borrower $B_i$ is pivotal with respect to the size of the group. The influence of each borrower to the lender $L$ is equal to the normalized value of the final intensity measure.

For the "many lenders, many borrowers" case the KBI is aggregated over all lenders taking into account the size of each lender's total loans.

For our hypothetical example suppose that the threshold value $q$ is equal to 25% of all outgoing links for each element. Then, using the above-mentioned methodology, we obtain the values of the intensities of interaction taking into account direct and indirect components (see Table 3).

**Table 3.** Key Borrower Index for Numerical Example 1

| Index\Agents | 1 | 2 | 3 | 4 | 5 | 6 | 7 | 8 | 9 | 10 |
|---|---|---|---|---|---|---|---|---|---|---|
| Key Borrower Index, q=25% | 0 | 0.152 | 0 | 0 | 0.121 | 0.356 | 0.019 | 0.019 | 0.212 | 0.121 |

A more detailed information on index calculation is provided in Appendix 1.

As a result, the KBI that takes into consideration only short-range interactions as well as classical centrality measures considers borrowers 6 and 9 as the most pivotal. The importance of the borrower 10 is still underestimated which can be explained by the following reasons.

First, in (Aleskerov et al., 2014) only direct interactions of the first level were taken into account, while we deal with the long-range links between elements of the network and analyze all three channels of influence, including the third, which has not considered yet.

Second, different centrality measures and KBI very poorly estimate the characteristics of the system elements in terms of stocks. In other words, they took into account only flows between borrowers and creditors disregarding their size.

Moreover, for the case of many lenders the idea of aggregation the KBI over all lenders does not take into account on how closely connected different lenders are. Moreover, it does not fully take into account chain reactions in the whole system.

Thus, the existing methods for assessing the impact of networks have three major drawbacks: 1) many of them do not take into account the intensity of the interaction of agents in networks; 2) they do not consider multistep interaction of agents; 3) they do not take into account the size of the vertex relative to those with whom they interact.

We have provided a simple example that allows visualizing all of these drawbacks. In Numerical Example 1, one can find a key vertex not captured by existing methods as a guesstimate. However, in the case of real financial and other types of networks detection such vertices approximately is not always possible.

We propose a new method for assessing of the agents' influence in the network, long-range interaction centrality (LRIC) (Aleskerov et. al., 2016). Our approach is based on a very simple observation. When we consider the network of interconnected lenders and borrowers, each creditor's sustainability will be affected by its direct borrowers. In addition, bankruptcy of any of direct borrowers may occur due to the bankruptcy of those ones to whom they were given loans, i.e. both direct and indirect borrowers are in relation to the original creditor.

In other words, our methodology allows us to consider the interaction between a lender and a borrower not just on the first level, but also on some levels beyond. For example, the agent 6 is not a direct borrower of the agent 1. However, it does not mean that he has no effect on him as it is assumed in the KBI (we can see it on Figure 1). Five channels can be identified here: a) through the agent 2; b) through the agent 3; c) through the agent 5; d) through the agent 2 via the agent 3, when we consider long-range influence; e) through the agent 5 via the agents 4 and 7, when we consider long-range influence.

There are two different ideas on how to take into account long-range interactions between members of the network. The first one is a distance-based approach where all different paths are considered for each member and somehow aggregated into a single value. The second one is based on the idea of simulations where we analyze the influence of individual members and their combinations to the whole network. Both ideas have an easy interpretation and can be applied to different areas.

The results of the proposed methods for our Numerical Example are shown in Table 4. A more detailed information on indices calculation is provided in Appendix 1.

**Table 4.** Long-range interaction centralities (LRIC) for Numerical Example 1

| Index\Agents | Long-range interaction (LRI) influence | | | | | | | | | |
|---|---|---|---|---|---|---|---|---|---|---|
| | **1** | **2** | **3** | **4** | **5** | **6** | **7** | **8** | **9** | **10** |
| LRI based on paths | 0 | 0.09 | 0 | 0 | 0.09 | **0.22** | 0.09 | 0.09 | **0.23** | **0.19** |
| LRI based on simulations | 0 | 0.085 | 0 | 0 | 0.085 | **0.211** | 0.072 | 0.071 | **0.216** | **0.261** |

As it is shown above, the results of our method differ from those that were obtained by classical centrality indices and the KBI. If long-range interactions are taken into account, agents 6, 9 and 10 will be considered as the most pivotal in the system. In our opinion, these results represent the actual power distribution in the network.

In the next Section we describe in details a new model for assessing the agents' influence in the network.

# 3. The model

The aim of this Section is to explain in details our approach and demonstrate how it works for the Numerical Example 2 (see Figure 2). As for Numerical Example 1, we consider a complex system of interconnections for agent's lending activities. The values on the edges as in previous Section represent the amount of loan that one agent gave to another one, and network structure corresponds to the bow-tie representation. Moreover, this network is structurally closer to the actually existing network of financial interactions, as links between the agents are more diversified, and comparing to the Numerical Example 1 there are less strongly connected components. As it was mentioned before, we propose two methods to find key pivotal borrowers in the system.

The main difference from the KBI is that our model takes into account $s$-long-range borrowers for each lender. In many problems interactions of indirect neighbors play a significant role in the whole system, hence, there is a need to consider these links. The parameter $s$ defines how many "layers" are examined for each lender, it depends on the problem and in general case can be unspecified so all possible direct and indirect neighbors are taken into account.

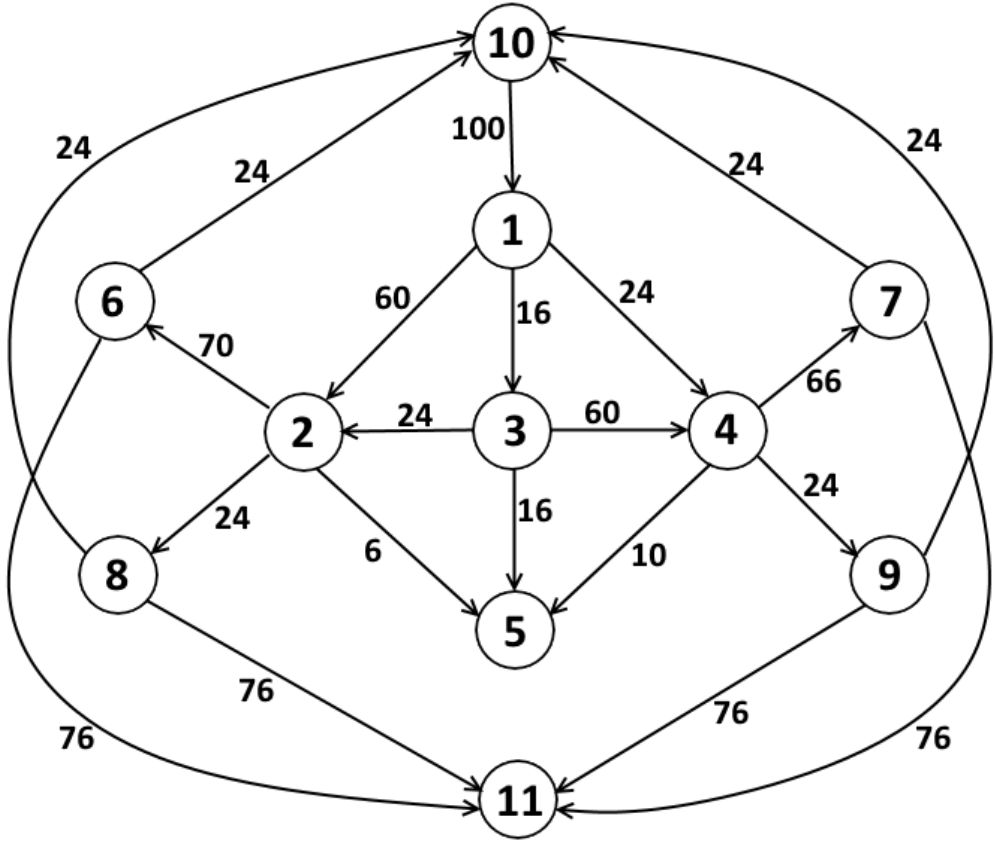

**Figure 2**. Numerical Example 2

To describe the proposed approach some definitions are given below.

Consider a set of members $N$, $N = \{1, \ldots, n\}$, and a matrix $A = [a_{ij}]$, where $i, j \in N$ and $a_{ij}$ is the loan from the member $i$ to the member $j$. For simplicity suppose the matrix $A$ being already transformed, i.e., if $a_{ij} \neq 0$ then $a_{ji} = 0$.

Denote by $N_i$ a set of direct neighbors of the $i$-th member, i.e., $N_i = \{j \in N: a_{ij} \neq 0\}$. Obviously, the total number of possible groups of direct neighbors for the member $i$ is equal to $2^{|N_i|}$.

*Definition 1.* The group of direct neighbors of the $i$-th member $\Omega(i) \subseteq N_i$ is critical if $\sum_{j \in \Omega(i)} a_{ij} \geq q_i$, where $q_i$ is a predefined threshold of the $i$-th member.

*Definition 2.* The member $x \in \Omega(i)$ is pivotal if $\sum_{j \in \Omega(i) \setminus \{x\}} a_{ij} < q_i$. Denote by $\Omega_p(i)$ a set of pivotal members in the group $\Omega(i)$, i.e., $\Omega_p(i) = \left\{ y \in \Omega(i) \mid \sum_{j \in \Omega(i) \setminus \{y\}} a_{ij} < q_i \right\}$.

For our Numerical Example 2 there are 11 agents in the system ($n$=11): 9 of them are both lenders and borrowers while 2 remaining elements are pure borrowers. So, we could form the matrix $A$ of the size of $11 \times 11$. The sets of direct neighbors $N_i$ and critical groups when $q_i = 25\%$ for each element are shown in Table 5.

**Table 5.** Direct borrowers and critical groups for Numerical Example 2

| Element, $i$ | Direct borrowers, $N_i$ | Critical groups, $\Omega(i)$, $q = 25\%$ |
|---|---|---|
| 1 | $\{2, 3, 4\}$ | $\{2\}, \{2, 3\}, \{2, 4\},$ $\{3, 4\}, \{2, 3, 4\},$ |
| 2 | $\{5, 6, 8\}$ | $\{6\}, \{5, 6\}, \{5, 8\},$ $\{6, 8\}, \{5, 6, 8\}$ |
| 3 | $\{2, 4, 5\}$ | $\{4\}, \{2, 4\}, \{2, 5\},$ $\{4, 5\}, \{2, 4, 5\}$ |
| 4 | $\{5, 7, 9\}$ | $\{7\}, \{5, 7\}, \{5, 9\},$ $\{7, 9\}, \{5, 7, 9\}$ |
| 5 | $\emptyset$ | $\emptyset$ |
| 6 | $\{10, 11\}$ | $\{11\}, \{10, 11\}$ |
| 7 | $\{10, 11\}$ | $\{11\}, \{10, 11\}$ |
| 8 | $\{10, 11\}$ | $\{11\}, \{10, 11\}$ |
| 9 | $\{10, 11\}$ | $\{11\}, \{10, 11\}$ |
| 10 | $\{1\}$ | $\{1\}$ |
| 11 | $\emptyset$ | $\emptyset$ |

Pivotal members for the element 1 when $q_1 = 25\%$ is provided in Table 6.

**Table 6.** Critical groups and pivotal member for the element 1

| Critical groups, $\Omega(1)$ | Pivotal members, $\Omega_p(1)$ |
|---|---|
| $\{2\}$ | $\{2\}$ |
| $\{2,3\}$ | $\{2\}$ |
| $\{2,4\}$ | $\{2\}$ |
| $\{3,4\}$ | $\{3, 4\}$ |
| $\{2,3,4\}$ | $\emptyset$ |

### 3.1. S-long-range interactions index based on paths

Let us construct a matrix $C = [c_{ij}]$ with respect to the matrix $A$ and predefined threshold as

$$c_{ij} = \begin{cases} \dfrac{a_{ij}}{\min\limits_{\Omega(i) \subseteq N_i \mid j \in \Omega_p(i)} \sum_{l \in \Omega(i)} a_{il}}, & if \ j \in \Omega_p(i) \subseteq N_i, \\ 0, & j \notin \Omega_p(i) \subseteq N_i, \end{cases}$$

where $\Omega(i)$ is a critical group of direct neighbors for the element $i$, $\Omega(i) \subseteq N_i$, and $\Omega_p(i)$ is pivotal group for the element $i$, $\Omega_p(i) \subseteq \Omega(i)$.

The construction of matrix $C$ is highly related to (Aleskerov et al., 2014) because it requires to consider separately each element of the system as a lender while other participants of the system are assumed as borrowers. The only difference here is that in our approach only groups of direct neighbors are considered.

The interpretation of matrix $C$ is rather simple. If $c_{ij} = 1$ then the borrower $j$ has a maximal influence to the lender $i$, i.e., the loan amount to the borrower $j$ is critical for the lender $i$. On the contrary, if $c_{ij} = 0$ then the borrower $j$ does not directly influence the lender $i$. Finally, the value $0 < c_{ij} < 1$ indicates the impact level of the borrower $j$ to the bankruptcy of the lender $i$.

Let us construct a matrix $C = [c_{ij}]$ for Numerical Example 2 with threshold value $q$=25% according to the approach based on paths. For example, when we want to estimate the direct influence of borrowers for the element 1, we search for a minimal critical group, i.e. a critical group with the lowest total loan from the element 1, where a particular borrower is pivotal and then estimate the direct influence $c_{1j}$. According to Table 4, the element 1 has 3 direct borrowers, hence, the minimal critical group for the element 2 is {2} and $c_{12} = \frac{60}{60} = 1$. For the elements 3 and 4 the minimal critical group is {3, 4} and $c_{13} = \frac{16}{16+24} = 0.4$ , $c_{14} = \frac{24}{16+24} = 0.6$. A graphical representation of the matrix is shown on Figure 3.

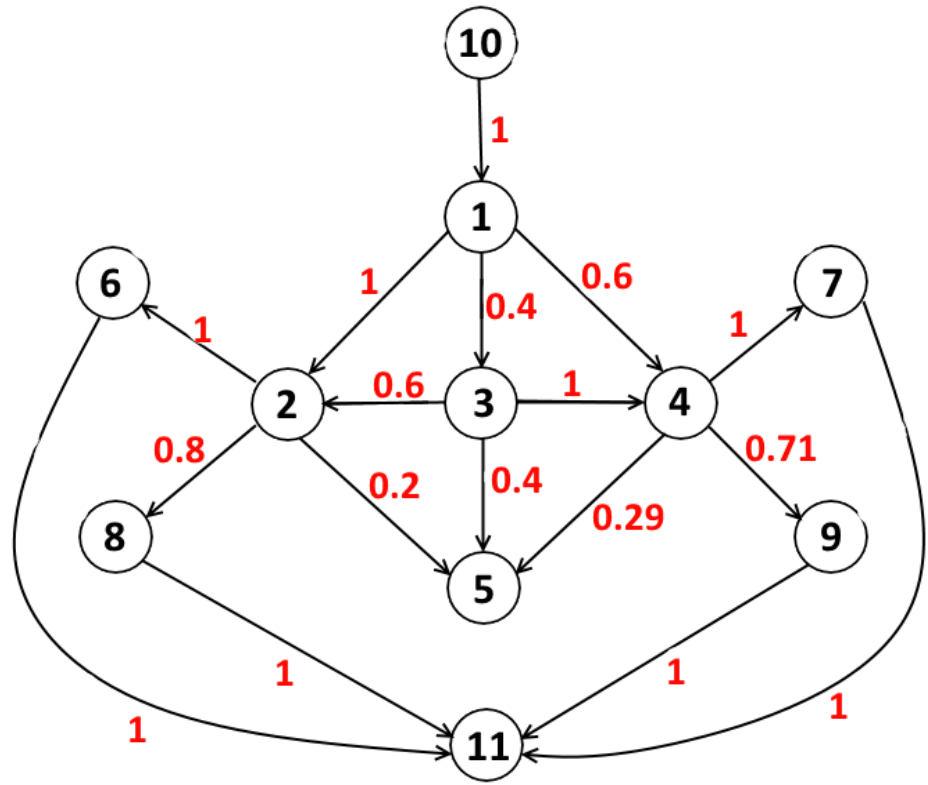

**Figure 3**. The network for matrix $C$.

In addition, we can see on Figure 3 that the element 10 does not influence any other element in the system.

Thus, we evaluated the direct influence of the first level of each element in the system. To define the indirect influence between two elements let us give a definition of the ρ-path.

Denote by $\rho$ a binary relation which is constructed as

$$i\rho j \Leftrightarrow c_{ij} > 0.$$

A pair ($i$, $j$) such that $i\rho j$ is called a ρ-step. A path from $i$ to $j$ is an ordered sequence of steps starting at $i$ and ending at $j$, such that the second element in each step coincides with the first element of the next step. If all steps in a path belong to the same relation $\rho$, we call it $\rho$-path, i.e., a $\rho$-path is an ordered sequence of elements $i$, $j_1$, ...,$j_k$, $j$, such that $i\rho j_1$, $j_1\rho j_2$, ..., $j_{k-1}\rho j_k$, $j_k\rho j$. The number of steps in a path is called the path's length.

To define the indirect influence between any two elements consider all $\rho$-paths between them of length less than some parameter $s$. Each path should not contain any cycles, i.e. there are no elements that occur in the $\rho$-path at least twice. For instance, there are only two paths between elements 8 and 1 from the Numerical Example 2 (see Figure 4): dashed lines, via element 2 ($8\rho2\rho1$), and dotted lines, via elements 3 and 2 ($8\rho2\rho3\rho1$).

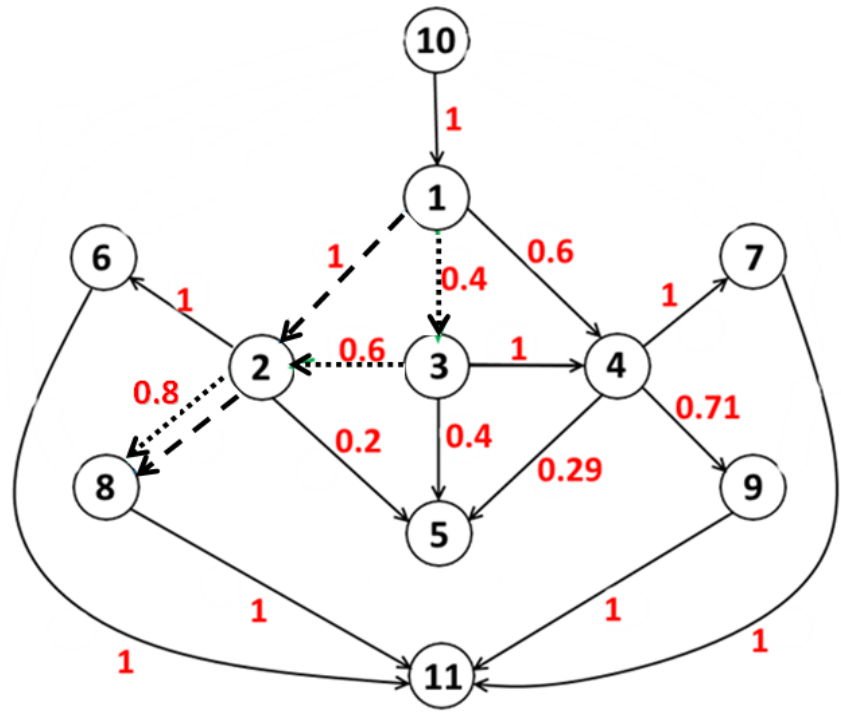

**Figure 4.** Paths between elements 8 and 1 (via element 2 - dashed lines, via elements 2 and 3 - dotted lines).

Denote by $P^{ij} = \{P_1^{ij}, P_2^{ij}, \ldots, P_m^{ij}\}$ a set of unique $\rho$-paths from $i$ to $j$, where $m$ is the total number of paths and denote by $n(k) = |P_k^{ij}|$, where $k = \overline{1, m}$, a length of the $k$-th path. Then we can define the indirect influence $f(P_k^{ij})$ between elements $i$ and $j$ via the k-th $\rho$-path $P_k^{ij}$ as

$$f(P_k^{ij}) = c_{ij(1,k)} \cdot c_{j(1,k)j(2,k)} \cdot \ldots \cdot c_{j(n(k),k)j}, \qquad (1)$$

or

$$f(P_k^{ij}) = \min(c_{ij(1,k)}, c_{j(1,k)j(2,k)}, \ldots, c_{j(n(k),k)j}), \qquad (2)$$

where $j(l, k)$, $l = \overline{1, n(k)}$ is the $l$-th element which occurs on $k$-th $\rho$-path from $i$ to $j$.

The interpretation of formulae (1) and (2) is the following. According to the formula (1) the total influence of the element $j$ to the element $i$ via the $k$-th $\rho$-path $P_k^{ij}$ is calculated as the aggregated value of direct influences between elements which are on the $k$-th $\rho$-path between $i$ and $j$ while the formula (2) defines the total influence as the minimal direct influence between any elements from the $k$-th $\rho$-path.

A simple example of indirect influence estimation between two elements is provided on Figure 5. In the first case the influence is proportional to the losses (risks) from bankruptcy of each borrower on the path while in the second case the influence is equal to the minimal risk of bankruptcy of the borrower which is on the path between elements 1 and 2.

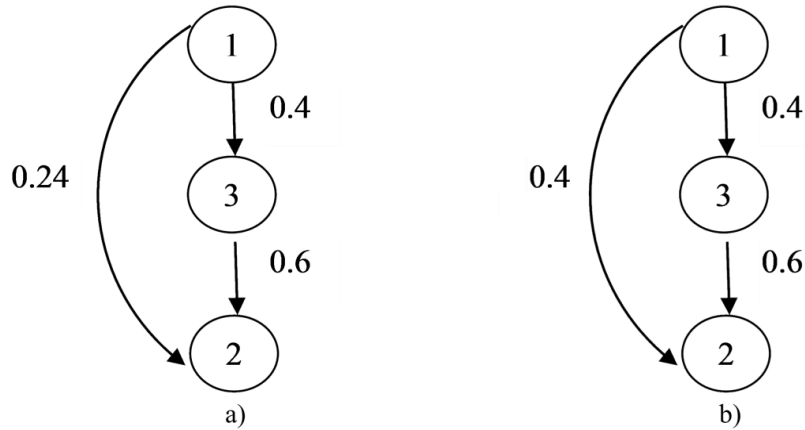

**Figure 5**. Indirect influence: a) multiplication of direct influences and
b) minimal direct influence

It is necessary to mention that in some cases there is no need to consider all possible paths between elements $i$ and $j$, i.e. we can assume that starting from some path's length $s$ indirect interactions does not influence the initial member. Thus, we designed the parameter $s$ that defines how many layers (path's length) are taken into account.

For example, consider all paths between elements 11 and 1 from Numerical Example 2 (see Figure 3). There are four paths between these elements: *11ρ8ρ2ρ1*, *11ρ9ρ4ρ1*, *11ρ8ρ2ρ3ρ1* and *11ρ9ρ4ρ3ρ1*. If parameter $s = 3$, so we will be interested only in first two paths, whereas the others will not be taken into account.

Since there can be many paths between two elements of the system, there is a problem of aggregating the influence of different paths. To estimate the aggregated indirect influence several methods are proposed.

The aggregated results will form a new matrix $C^*(s) = [c_{ij}^*(s)]$.

1. *The indirect influence: sum of paths influences*

$$c_{ij}^*(s) = \min(1, \sum_{k: n(k) \leq s} f(P_k^{ij})). \qquad (3)$$

2. *The indirect influence: maximal path influence*

$$c_{ij}^*(s) = \max_{k: n(k) \leq s} f(P_k^{ij}). \qquad (4)$$

3. *The indirect influence: the threshold rule*

The threshold aggregation was proposed in (Aleskerov et al., 2007) and the idea of the rule is rather simple. Suppose we have a set of elements and each element is evaluated by *n* grades that may have *m* different values. Then we can calculate for each element the values $v_1(k), v_2(k),\ldots, v_m(k)$ which contain information on how many *i*-th ($i = \overline{1, m}$) grades each element received. Then according to the threshold rule the element *x* *V*-dominates the element *y* if $v_1(x) < v_1(y)$ or, if there exists $d \leq m$, such that $v_h(x) = v_h(y), \; \forall h = 1, \ldots, d - 1$, and $v_d(x) < v_d(y)$. In other words, first, the number of worst places are compared, if these numbers are equal than the number of second worst places are compared, and so on. The element which is not dominated by any other element via *V* is considered as the best one.

Considering the threshold rule as one of possible ways on how the indirect influence can be evaluated, we propose the following aggregation procedure

$$c_{ij}^*(s) = f(P_z^{ij}), \qquad (5)$$

where

$$z = \operatorname{argmin}_{k: n(k) \leq s} v(P_k^{ij}), \qquad (6)$$

and

$$v(P_k^{ij}) = \sum_{l=1}^m v_1(P_k^{ij}) * (s + 1)^{m-l} + s - n(k).$$

The formula (6) is identical to the threshold rule (Aleskerov et al., 2010).

Note that if there is no path between elements *i* and *j*, then $c_{ij}^*(s) = 0$.

For example, consider that paths between elements 11 and 1. According to the formula (1) the influence of the path *11ρ8ρ2ρ1* (path's length = 3) is equal to 0.8, the influence of the path *11ρ8ρ2ρ3ρ1* (path's length = 4) is equal to 0.56, the influence of the path *11ρ9ρ4ρ1* (path's length = 3) is equal to 0.426, the influence of the path *11ρ9ρ4ρ3ρ1* (path's length = 4) is equal to 0.284.

If $s=3$, only two paths are taken into account. Thus, the total influence of the element 11 to the element 1 will be equal to 1 according to the sum of paths influences ($\min\{1, (0.8 + 0.56)\}$) or will be equal to 0.8 according to the maximal path influence ($\max\{0.8, 0.56\}$). For the threshold rule calculations are not so obvious since they depend on the system of grades which will be described later.

The interpretation of formulae (3)-(5) in term of borrowers is the following. The sum of paths influences is equivalent to the most pessimistic case of the indirect influence where we take into account all possible channels of risk from a particular borrower to the creditor. The maximal path influence and the influence assessment by the threshold rule help us to find the most vulnerable risk transmission channel.

Thus, we can define the indirect influence between elements $i$ and $j$ via all possible paths between these elements. The paths influences can be evaluated by formulae (1)-(2) and aggregated into a single value by formulae (3)-(5). Thus, 6 combinations are possible for matrix $C^*(s)$ construction (see Table 7). In our opinion, all possible combinations of formulae have a sense except the combination of formulae (2) and (3).

**Table 7.** Possible combinations of methods for indirect influence

| | | Paths aggregation | | |
| --- | --- | --- | --- | --- |
| | | Sum of paths influences | Maximal path influence | Threshold rule |
| **Path influence** | **Multiplication of direct influence** | SumPaths | MaxPath | MultT |
| | **Minimal direct influence** | – | MaxMin | MaxT |

For our Numerical Example 2 we can differently construct the matrix $C^*$ which represents the total influence and is used for aggregation of influences into a single vector with respect to the weights.

For instance, let us calculate the influence of the element 5 to the element 1. Consider all possible paths from the element 5 to the element 1. They are shown in Table 8.

**Table 8.** Possible paths between elements 5 and 1

| ID | Path | Multiplication of path influences | Minimal direct influence |
|----|------|-----------------------------------|--------------------------|
| 1 | $5 \underset{0.2}{\to} 2 \underset{1}{\to} 1$ | 0.2 | 0.2 |
| 2 | $5 \underset{0.4}{\to} 3 \underset{0.4}{\to} 1$ | 0.16 | 0.4 |
| 3 | $5 \underset{0.29}{\to} 4 \underset{0.6}{\to} 1$ | 0.174 | 0.29 |
| 4 | $5 \underset{0.2}{\to} 2 \underset{0.6}{\to} 3 \underset{0.4}{\to} 1$ | 0.048 | 0.2 |
| 5 | $5 \underset{0.29}{\to} 4 \underset{1}{\to} 3 \underset{0.4}{\to} 1$ | 0.116 | 0.29 |

Thus, there are five possible ways how the element 5 influences the element 1. Let us now aggregate this information into a single value by different methods. To compare different paths by the threshold rule the following grades of direct influence were developed.

*Grades:*

0. $c_{ij} = 0$;
1. $0 < c_{ij} \le 0.25$;
2. $0.25 < c_{ij} \le 0.5$;
3. $0.5 < c_{ij} \le 0.8$;
4. $0.8 < c_{ij} \le 1$.

Now we can define the path between the elements 5 and 1 according to the threshold rule. Note that for the threshold rule the values on the edges are equal to the grades which was proposed above. The results are provided in Table 9.

**Table 9.** Paths aggregation by the threshold rule, *s=3*

| ID, $k$ | Path | Path (grades on edges) | Paths influence, $v\left(P_k^{15}\right)$ |
|---------|------|------------------------|-------------------------------------------|
| 1 | $5 \underset{0.2}{\to} 2 \underset{1}{\to} 1$ | $5 \underset{1}{\to} 2 \underset{4}{\to} 1$ | 66 |
| 2 | $5 \underset{0.4}{\to} 3 \underset{0.4}{\to} 1$ | $5 \underset{2}{\to} 3 \underset{2}{\to} 1$ | 33 |
| 3 | $5 \underset{0.29}{\to} 4 \underset{0.6}{\to} 1$ | $5 \underset{2}{\to} 4 \underset{3}{\to} 1$ | 21[*] |
| 4 | $5 \underset{0.2}{\to} 2 \underset{0.6}{\to} 3 \underset{0.4}{\to} 1$ | $5 \underset{1}{\to} 2 \underset{3}{\to} 3 \underset{2}{\to} 1$ | 84 |
| 5 | $5 \underset{0.29}{\to} 4 \underset{1}{\to} 3 \underset{0.4}{\to} 1$ | $5 \underset{2}{\to} 4 \underset{4}{\to} 3 \underset{2}{\to} 1$ | 33 |

[*]the path chosen by the threshold rule

The overall results are provided in Table 10.

**Table 10.** The total influence of the element 5 to the element 1 by different methods

| Method | Considered paths IDs | Influence |
|---|---|---|
| SumPaths | 1-5 | 0.698 |
| MaxPath | 1 | 0.2 |
| MaxMin | 2 | 0.4 |
| MultT | 3 | 0.174 |
| MaxT | 3 | 0.29 |

Similarly, we can estimate the influence of any other elements and construct the matrix $C^*$ according to different methods. The results are provided in Table 11-15.

**Table 11.** Matrix $C^*$ for the Numerical Example 2, SumPaths

| | 1 | 2 | 3 | 4 | 5 | 6 | 7 | 8 | 9 | 10 | 11 | Weights |
|---|---|---|---|---|---|---|---|---|---|---|---|---|
| **1** | 0 | 1 | 0.40 | 1 | 0.70 | 1 | 1 | 0.99 | 0.71 | 0 | 1 | 0.11 |
| **2** | 0 | 0 | 0 | 0 | 0.20 | 1 | 0 | 0.80 | 0 | 0 | 1 | 0.11 |
| **3** | 0 | 0.60 | 0 | 1 | 0.81 | 0.60 | 1 | 0.48 | 0.71 | 0 | 1 | 0.11 |
| **4** | 0 | 0 | 0 | 0 | 0.29 | 0 | 1 | 0 | 0.71 | 0 | 1 | 0.11 |
| **5** | 0 | 0 | 0 | 0 | 0 | 0 | 0 | 0 | 0 | 0 | 0 | 0 |
| **6** | 0 | 0 | 0 | 0 | 0 | 0 | 0 | 0 | 0 | 0 | 1 | 0.11 |
| **7** | 0 | 0 | 0 | 0 | 0 | 0 | 0 | 0 | 0 | 0 | 1 | 0.11 |
| **8** | 0 | 0 | 0 | 0 | 0 | 0 | 0 | 0 | 0 | 0 | 1 | 0.11 |
| **9** | 0 | 0 | 0 | 0 | 0 | 0 | 0 | 0 | 0 | 0 | 1 | 0.11 |
| **10** | 1 | 1 | 0.40 | 1 | 0.70 | 1 | 1 | 0.99 | 0.71 | 0 | 1 | 0.11 |
| **11** | 0 | 0 | 0 | 0 | 0 | 0 | 0 | 0 | 0 | 0 | 0 | 0 |
| **Total** | 0.11 | 0.29 | 0.09 | 0.33 | 0.30 | 0.40 | 0.44 | 0.36 | 0.31 | 0 | 1 | |
| **Total (normalized)** | **0.03** | **0.08** | **0.02** | **0.09** | **0.08** | **0.11** | **0.12** | **0.10** | **0.09** | **0** | **0.27** | |

**Table 12.** Matrix $C^*$ for the Numerical Example 2, MaxPath

| | 1 | 2 | 3 | 4 | 5 | 6 | 7 | 8 | 9 | 10 | 11 | Weights |
|---|---|---|---|---|---|---|---|---|---|---|---|---|
| **1** | 0 | 1 | 0.40 | 0.60 | 0.20 | 1 | 0.60 | 0.80 | 0.42 | 0 | 1 | 0.11 |
| **2** | 0 | 0 | 0 | 0 | 0.20 | 1 | 0 | 0.80 | 0 | 0 | 1 | 0.11 |
| **3** | 0 | 0.60 | 0 | 1 | 0.40 | 0.60 | 1 | 0.48 | 0.71 | 0 | 1 | 0.11 |
| **4** | 0 | 0 | 0 | 0 | 0.29 | 0 | 1 | 0 | 0.71 | 0 | 1 | 0.11 |
| **5** | 0 | 0 | 0 | 0 | 0 | 0 | 0 | 0 | 0 | 0 | 0 | 0 |
| **6** | 0 | 0 | 0 | 0 | 0 | 0 | 0 | 0 | 0 | 0 | 1 | 0.11 |
| **7** | 0 | 0 | 0 | 0 | 0 | 0 | 0 | 0 | 0 | 0 | 1 | 0.11 |
| **8** | 0 | 0 | 0 | 0 | 0 | 0 | 0 | 0 | 0 | 0 | 1 | 0.11 |
| **9** | 0 | 0 | 0 | 0 | 0 | 0 | 0 | 0 | 0 | 0 | 1 | 0.11 |
| **10** | 1 | 1 | 0.40 | 0.60 | 0.20 | 1 | 0.60 | 0.80 | 0.42 | 0 | 1 | 0.11 |
| **11** | 0 | 0 | 0 | 0 | 0 | 0 | 0 | 0 | 0 | 0 | 0 | 0 |
| **Total** | 0.11 | 0.29 | 0.09 | 0.24 | 0.14 | 0.40 | 0.36 | 0.32 | 0.25 | 0 | 1 | |
| **Total (normalized)** | **0.03** | **0.09** | **0.03** | **0.08** | **0.04** | **0.12** | **0.11** | **0.10** | **0.08** | **0** | **0.31** | |

Table 13. Matrix $C^*$ for the Numerical Example 2, MaxMin

| | 1 | 2 | 3 | 4 | 5 | 6 | 7 | 8 | 9 | 10 | 11 | Weights |
|---|---|---|---|---|---|---|---|---|---|---|---|---|
| **1** | 0 | 1 | 0.40 | 0.60 | 0.40 | 1 | 0.60 | 0.80 | 0.60 | 0 | 1 | 0.11 |
| **2** | 0 | 0 | 0 | 0 | 0.20 | 1 | 0 | 0.80 | 0 | 0 | 1 | 0.11 |
| **3** | 0 | 0.60 | 0 | 1 | 0.40 | 0.60 | 1 | 0.60 | 0.71 | 0 | 1 | 0.11 |
| **4** | 0 | 0 | 0 | 0 | 0.29 | 0 | 1 | 0 | 0.71 | 0 | 1 | 0.11 |
| **5** | 0 | 0 | 0 | 0 | 0 | 0 | 0 | 0 | 0 | 0 | 0 | 0 |
| **6** | 0 | 0 | 0 | 0 | 0 | 0 | 0 | 0 | 0 | 0 | 1 | 0.11 |
| **7** | 0 | 0 | 0 | 0 | 0 | 0 | 0 | 0 | 0 | 0 | 1 | 0.11 |
| **8** | 0 | 0 | 0 | 0 | 0 | 0 | 0 | 0 | 0 | 0 | 1 | 0.11 |
| **9** | 0 | 0 | 0 | 0 | 0 | 0 | 0 | 0 | 0 | 0 | 1 | 0.11 |
| **10** | 1 | 1 | 0.40 | 0.60 | 0.40 | 1 | 0.60 | 0.80 | 0.60 | 0 | 1 | 0.11 |
| **11** | 0 | 0 | 0 | 0 | 0 | 0 | 0 | 0 | 0 | 0 | 0 | 0 |
| **Total** | 0.11 | 0.29 | 0.09 | 0.24 | 0.19 | 0.40 | 0.36 | 0.33 | 0.29 | 0 | 1 | |
| **Total (normalized)** | **0.03** | **0.09** | **0.03** | **0.07** | **0.06** | **0.12** | **0.11** | **0.10** | **0.09** | **0** | **0.30** | |

Table 14. Matrix $C^*$ for the Numerical Example 2, MultT

| | 1 | 2 | 3 | 4 | 5 | 6 | 7 | 8 | 9 | 10 | 11 | Weights |
|---|---|---|---|---|---|---|---|---|---|---|---|---|
| **1** | 0 | 1 | 0.40 | 0.60 | 0.18 | 1 | 0.60 | 0.80 | 0.42 | 0 | 1 | 0.11 |
| **2** | 0 | 0 | 0 | 0 | 0.20 | 1 | 0 | 0.80 | 0 | 0 | 1 | 0.11 |
| **3** | 0 | 0.60 | 0 | 1 | 0.40 | 0.60 | 1 | 0.48 | 0.71 | 0 | 1 | 0.11 |
| **4** | 0 | 0 | 0 | 0 | 0.29 | 0 | 1 | 0 | 0.71 | 0 | 1 | 0.11 |
| **5** | 0 | 0 | 0 | 0 | 0 | 0 | 0 | 0 | 0 | 0 | 0 | 0 |
| **6** | 0 | 0 | 0 | 0 | 0 | 0 | 0 | 0 | 0 | 0 | 1 | 0.11 |
| **7** | 0 | 0 | 0 | 0 | 0 | 0 | 0 | 0 | 0 | 0 | 1 | 0.11 |
| **8** | 0 | 0 | 0 | 0 | 0 | 0 | 0 | 0 | 0 | 0 | 1 | 0.11 |
| **9** | 0 | 0 | 0 | 0 | 0 | 0 | 0 | 0 | 0 | 0 | 1 | 0.11 |
| **10** | 1 | 1 | 0.40 | 0.60 | 0.18 | 1 | 0.60 | 0.80 | 0.42 | 0 | 1 | 0.11 |
| **11** | 0 | 0 | 0 | 0 | 0 | 0 | 0 | 0 | 0 | 0 | 0 | 0 |
| **Total** | 0.11 | 0.29 | 0.09 | 0.24 | 0.14 | 0.40 | 0.36 | 0.32 | 0.25 | 0 | 1 | |
| **Total (normalized)** | **0.03** | **0.09** | **0.03** | **0.07** | **0.04** | **0.12** | **0.11** | **0.10** | **0.08** | **0** | **0.31** | |

Table 15. Matrix $C^*$ for the Numerical Example 2, MaxT

| | 1 | 2 | 3 | 4 | 5 | 6 | 7 | 8 | 9 | 10 | 11 | Weights |
|---|---|---|---|---|---|---|---|---|---|---|---|---|
| **1** | 0 | 1 | 0.40 | 0.60 | 0.29 | 1 | 0.60 | 0.80 | 0.60 | 0 | 1 | 0.11 |
| **2** | 0 | 0 | 0 | 0 | 0.20 | 1 | 0 | 0.80 | 0 | 0 | 1 | 0.11 |
| **3** | 0 | 0.60 | 0 | 1 | 0.40 | 0.60 | 1 | 0.60 | 0.71 | 0 | 1 | 0.11 |
| **4** | 0 | 0 | 0 | 0 | 0.29 | 0 | 1 | 0 | 0.71 | 0 | 1 | 0.11 |
| **5** | 0 | 0 | 0 | 0 | 0 | 0 | 0 | 0 | 0 | 0 | 0 | 0 |
| **6** | 0 | 0 | 0 | 0 | 0 | 0 | 0 | 0 | 0 | 0 | 1 | 0.11 |
| **7** | 0 | 0 | 0 | 0 | 0 | 0 | 0 | 0 | 0 | 0 | 1 | 0.11 |
| **8** | 0 | 0 | 0 | 0 | 0 | 0 | 0 | 0 | 0 | 0 | 1 | 0.11 |
| **9** | 0 | 0 | 0 | 0 | 0 | 0 | 0 | 0 | 0 | 0 | 1 | 0.11 |
| **10** | 1 | 1 | 0.40 | 0.60 | 0.29 | 1 | 0.60 | 0.80 | 0.60 | 0 | 1 | 0.11 |
| **11** | 0 | 0 | 0 | 0 | 0 | 0 | 0 | 0 | 0 | 0 | 0 | 0 |
| **Total** | 0.11 | 0.29 | 0.09 | 0.24 | 0.16 | 0.40 | 0.36 | 0.33 | 0.29 | 0 | 1 | |
| **Total (normalized)** | **0.03** | **0.09** | **0.03** | **0.08** | **0.05** | **0.13** | **0.11** | **0.10** | **0.09** | **0** | **0.31** | |

The aggregation of matrix $C^*(s)$ into a single vector that shows the total influence of each element of the system can be done with respect to the weights (importance) of each element as it is done in (Aleskerov et al., 2014).

As a result, we can see that elements 6, 7 and 11 are considered as the most pivotal in the system while the influence of the element 5 is more than the influence of the elements 1 and 10. Now let us consider the second approach based on simulations.

### 3.2. S-Long-Range Interactions Centrality index based on simulations

Another approach of estimating the power of each element in the system is based on the following idea. Suppose that some borrowers are not capable to return the loan. Will they form critical group? Will it lead to the fact that their creditors in turn will not cover the loans to other creditors?

More formally, let a construct a matrix $C = [c_{ij}]$ with respect to the matrix $A$ and predefined threshold as

$$c_{ij} = \begin{cases} 1, if\ a_{ij} \geq q_i \\ \dfrac{a_{ij}}{q_i}, if\ 0 < a_{ij} < q_i \\ 0, if\ a_{ij} = 0. \end{cases}$$

In other words, the matrix $C$ indicates what share of the threshold value (critical loan amount) the element $i$ gave to the element $j$. The matrix $C$ is used to evaluate the long-range influence between elements of the system through simulations. A graphical representation of the matrix is shown on Figure 6.

Similarly, let us say that the group $\Omega(i) \subseteq N_i$ is critical if $\sum_{k \in \Omega(i)} c_{ik} \geq 1$, and any element in group $\Omega(i)$ is not capable to return the loan. If the group $\Omega(i)$ exists, then $i$ is not capable to return the loan to his own creditor.

Assume now that some borrowers (total number is $k_0$) are not capable to return a loan. Then, we can define a list of borrowers (total number is $k_1$) for which $k_0$ borrowers form a critical groups. Similarly, we can define a list of borrowers (total number is $k_2$) for which $k_0 + k_1$ borrowers form the critical group. The procedure continues until the predefined limit in the number of stages $s$ is reached or there is a stage $r$ (in the worst case when the parameter $s$ is undefined $r$ is less than or equal the diameter of the network) such that $k_r = 0$. Thus, we derived a list of borrowers (total number is $\sum_{l=1}^{\min(r,s)} k_l$) that are not capable to return their loans if we assume that $k_0$ borrowers cannot return their loans.

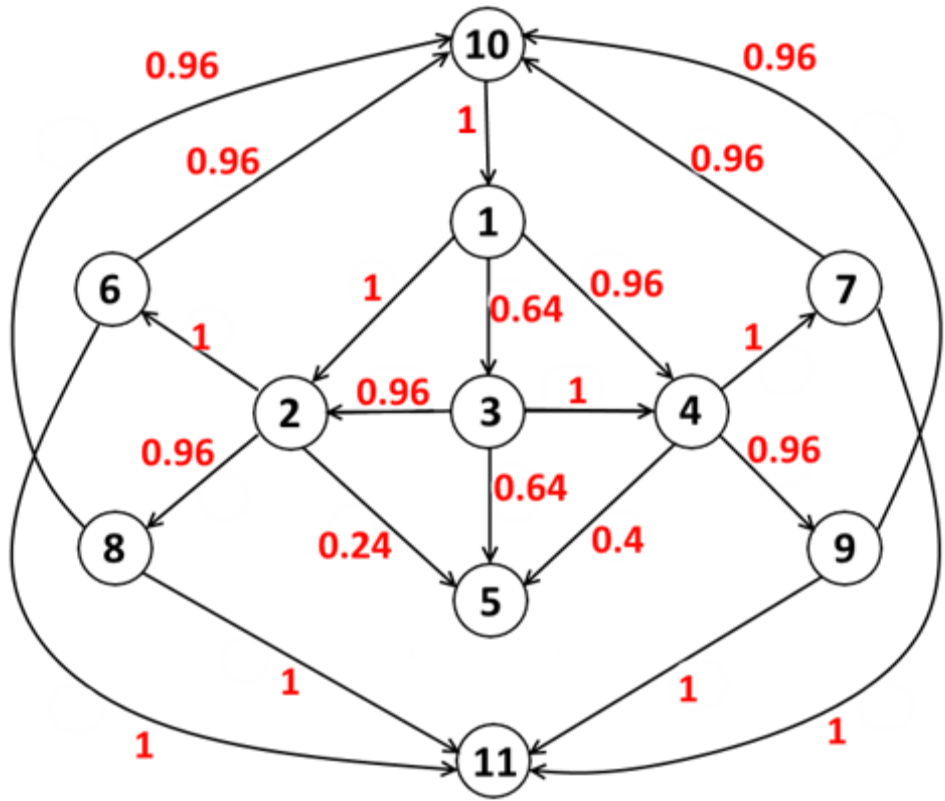

**Figure 6**. A graphical representation of matrix *C* for simulations approach.

For instance, suppose that elements 5, 6, 9 are not capable to return their loans. Then we can define which elements in turn will not be able to cover their loans (see Table 16).

**Table 16.** Simulation procedure for the combination $\{5, 6, 9\}$

| Step | Bankrupted Elements | Total number of elements |
|------|---------------------|--------------------------|
| 0 | $\{5, 6, 9\}$ | $k_0=3$ |
| 1 | $\{2, 4\}$ | $k_1=2$ |
| 2 | $\{1, 3\}$ | $k_2=2$ |
| 3 | $\{10\}$ | $k_3=1$ |

Thus, the bankruptcy of elements 5, 6 and 9 will lead to the bankruptcy of other five elements.

Similarly, we can assume any other combination of borrowers that cannot return their loans and define a list of all bankrupted borrowers. The results will form a new matrix $C^*(s) = [c_{ij}^*(s)]$, which shows in what percentage of cases the borrower $i$ could not return its loans if we assume that the borrower $j$ is not capable to return his own loans.

The interpretation of the matrix $C^*(s)$ is rather simple. If the value $c_{ij}^*$ is close to 1, then borrower $j$ is very critical for the borrower $i$. On the contrary, if the value $c_{ij}^*$ is close to 0, then the borrower $j$ hardly influences the borrower $i$. The results can be aggregated into a single vector that shows the total influence of each element in the system.

Thus, we can construct the matrix $C^*(s) = [c_{ij}^*(s)]$ for the Numerical Example 2 (see Table 17).

**Table 17.** Matrix $C^*$ for the Numerical Example 2 (simulations)

| | 1 | 2 | 3 | 4 | 5 | 6 | 7 | 8 | 9 | 10 | 11 | Weights |
|---|---|---|---|---|---|---|---|---|---|---|---|---|
| **1** | 0 | 1 | 0 | 1 | 0.59 | 1 | 1 | 0.43 | 0.42 | 0 | 1 | 0.11 |
| **2** | 0 | 0 | 0 | 0 | 0.41 | 1 | 0 | 0.42 | 0 | 0 | 1 | 0.11 |
| **3** | 0 | 0.57 | 0 | 1 | 0.92 | 0.48 | 1 | 0.47 | 0.45 | 0 | 1 | 0.11 |
| **4** | 0 | 0 | 0 | 0 | 0.40 | 0 | 1 | 0 | 0.42 | 0 | 1 | 0.11 |
| **5** | 0 | 0 | 0 | 0 | 0 | 0 | 0 | 0 | 0 | 0 | 0 | 0.00 |
| **6** | 0 | 0 | 0 | 0 | 0 | 0 | 0 | 0 | 0 | 0 | 1 | 0.11 |
| **7** | 0 | 0 | 0 | 0 | 0 | 0 | 0 | 0 | 0 | 0 | 1 | 0.11 |
| **8** | 0 | 0 | 0 | 0 | 0 | 0 | 0 | 0 | 0 | 0 | 1 | 0.11 |
| **9** | 0 | 0 | 0 | 0 | 0 | 0 | 0 | 0 | 0 | 0 | 1 | 0.11 |
| **10** | 1 | 1 | 0 | 1 | 0.62 | 1 | 1 | 0.46 | 0.47 | 0 | 1 | 0.11 |
| **11** | 0 | 0 | 0 | 0 | 0 | 0 | 0 | 0 | 0 | 0 | 0 | 0.00 |
| **Total** | 0.11 | 0.29 | 0.00 | 0.33 | 0.33 | 0.39 | 0.44 | 0.20 | 0.20 | 0.00 | 1.00 | |
| **Total (normalized)** | **0.03** | **0.09** | **0** | **0.10** | **0.10** | **0.12** | **0.14** | **0.06** | **0.06** | **0** | **0.30** | |

One of the key advantages of this approach is that it accurately takes into account all chain reactions of the system, so-called domino or contagion effect.

One of the key questions of this approach is which borrowers should be chosen on the first stage. For the general case, we can choose each borrower the same number of times. However, in real-life problems this is not always the case since different elements have different probability of default. It means that these probabilities can be taken into account at the simulation stage.

Another important issue of this approach is how we should define pivotal borrowers for each lender. Obviously, it is possible that for the lender $i$ there is a group of borrowers of size $k^* << k_0$ that will form a critical group and, consequently, lead to bankruptcy of the lender $i$. In other words, not all $k_0$ borrowers equally influence a specific lender. Thus, there is a need to find a minimal set of pivotal borrowers $\Omega_p(i)$ for each lender $i$ on each simulation stage. Currently, there is no solution for this problem except considering all possible subsets from the set of $k_0$ elements, which is one of the drawbacks of our approach.

A list of pivotal borrowers for each bankrupted element due to the bankruptcy of elements 5, 6 and 9 is provided in Table 18.

**Table 18.** Key borrowers detection for simulation procedure for the combination {5, 6, 9}

| Borrower $i$ | Critical groups, $\Omega(i)$ | Pivotal borrowers from {5, 6, 9}, $\Omega_p(i)$ |
|---|---|---|
| {1} | {2}, {2, 3}, {2, 4}, {3, 4}, {2, 3, 4} | {5, 6, 9} |
| {2} | {6}, {5, 6} | {6} |
| {3} | {4}, {2, 4}, {2, 5}, {4, 5}, {2, 4, 5} | {5, 6, 9} |
| {4} | {5, 9} | {5, 9} |
| {10} | {1} | {5, 6, 9} |

Another drawback of the proposed idea is the high computational complexity since on the simulation stage we should consider a large amount of combinations of borrowers, which are assumed of not being capable to return their loans. It leads to the fact that the value $k_0$ should be somehow constrained which actually sounds reasonable since the probability that large number of borrowers will not be able to return their loans at the same time is very small. One of the solutions that allow to decrease the computational complexity is to set some limits on the number of combinations, chain reactions or to add some probabilities of bankruptcy of each borrower.

Finally, we try to compare the results according to our model with centrality measures and the Key Borrower Index proposed in (Aleskerov et al., 2014).

Let us calculate the centrality measures and the Key Borrower Index. The results are shown in Table 19. More information on calculations of these indices is provided in Appendix 2.

**Table 19.** Centrality measures and Key Borrower Index for Numerical Example 2

| Indices\Agents | | 1 | 2 | 3 | 4 | 5 | 6 | 7 | 8 | 9 | 10 | 11 |
|---|---|---|---|---|---|---|---|---|---|---|---|---|
| Centrality measures | Weighted In-degree | **100** | 84 | 16 | 84 | 32 | 70 | 66 | 24 | 24 | **96** | **304** |
| | Weighted Out-degree | 100 | 100 | 100 | 100 | 0 | 100 | 100 | 100 | 100 | 100 | 0 |
| | Weighted Degree Difference | **0** | 16 | 84 | 16 | **-32** | 30 | 34 | 76 | 76 | 4 | **-304** |
| | Weighted Degree | **200** | 184 | 116 | 184 | 32 | 170 | 166 | 124 | 124 | **196** | **304** |
| | Closeness in | **0.0014** | **0.0012** | 0.001 | 0.001 | 0.001 | 0.001 | 0.001 | 0.001 | 0.001 | **0.0012** | **0.002** |
| | Closeness out | 0.001 | 0.0007 | 0.001 | 0.001 | **0.009** | **0.007** | 0.001 | 0.001 | 0.001 | 0.0011 | **0.009** |
| | Betweenness | **45** | **23** | 0 | 17 | 0 | 10 | 10 | 0 | 0 | **43** | 0 |
| | Eigenvector | 0.61 | 0.57 | 0.28 | 0.47 | 0.07 | **0.7** | **0.65** | 0.56 | 0.55 | 0.64 | **1** |
| | PageRank | **0.11** | **0.10** | 0.05 | 0.08 | 0.05 | 0.095 | 0.08 | 0.06 | 0.05 | 0.09 | **0.22** |
| Key Borrower Index, q=25% | | **0.11** | 0.10 | 0.01 | **0.11** | 0.01 | 0.10 | 0.10 | 0.01 | 0.01 | 0 | **0.44** |

As it is shown above, most centrality measures as well as the Key Borrower Index consider elements 1 and 11 as pivotal. The influence of the element 1 can be explained by the fact that it directly influences the element 10, which is highly interconnected with other elements of the system. However, if we accurately analyze the element 10, we will see that this borrower does not influence any lender since his loan is less than the critical threshold value for each lender and this element is not pivotal in any groups of borrowers. Thus, the influence of the element 1 is overestimated by all indices while the influence of the element 10 should be equal to zero. On the contrary, the influence of the element 5 is underestimated; however, this element directly and indirectly influences elements 1, 2, 3, 4 and 10. Another important issue is that none of these indices except eigenvector centrality considers elements 6 and 7 as pivotal. However, these elements directly influence elements 2 and 4 and indirectly influence elements 1,3,10.

Both our proposed approaches take into account these observations and consider elements 6, 7 and 11 as the most pivotal in the system. The results are similar to the results of the eigenvector centrality, however, the eigenvector centrality also highly evaluates the power of elements 1 and 10 while according to our methods their influence is rather small (it is 0 for the element 10).

It is also important to note that the key element does not necessarily should be at the end of the chain, i.e. play a role of pure borrower. The most pivotal element can also be located in the center of the network.

# 3. Empirical application - country assessment

In this Section the model outlined so far will be applied for evaluation of the level of banking systems interconnectedness. We try to detect countries with the most interconnected financial systems taking into consideration the intensities of countries' banking systems interactions. At the same time, we understand the limitations of the existing data. The analysis of cross-border country exposures relies primarily on data aggregated at the level of countries, and, hence, overlooks bank-level heterogeneity.

The data is taken from the Bank of International Settlements (BIS) statistics[f]. More precisely, we use the BIS consolidated banking statistics on an ultimate risk basis. For example, suppose that a bank from country A extends a loan to a company from country B and the loan is guaranteed by a bank from country C. On an ultimate risk basis, this loan would be reported as a claim on the country C because, if the company from B were unable to meet its obligations, then ultimately the bank from A would be exposed to the bank from C that guaranteed the loan. In other words, claims are allocated to the country where the final risk lies.

Foreign claims in CBS statistics are designed to analyze the exposure of internationally active banks to individual countries and sectors. The sectoral classification consists of a) banks;

---

b) official sector, which includes general government[g], central banks and international organizations; c) non-bank private sector, including non-bank financials.

Thus, our figures indicate the *i*-th banking system foreign claims on borrowers from different sectors in country *j*, which include its worldwide consolidated direct cross-border claims on country *j* plus the positions booked by its affiliates (subsidiaries and branches) in country *j* vis-à-vis residents of country *j*. The data covers on-balance sheet claims as well as some off-balance sheet exposures of banks headquartered in the reporting country and provides a measure of country credit risk exposures consonant with banks' own risk management systems.

The reporting countries comprise the G10 countries (Belgium, Canada, France, Germany, Japan, Netherlands, Sweden, Switzerland, United Kingdom, and USA) plus Australia, Austria, Chile, Finland, Greece, India, Ireland, South Korea, Portugal, Spain and Turkey.

BIS consolidated banking statistics apart from information about banking foreign claims includes aggregated data on regional country groupings such as regional residuals of developed countries, offshore centers, Africa and the Middle East, Asia-Pacific, Europe, Latin America and the Caribbean and "Unallocated" claims. These positions are used in BIS Statistics in cases when the balance of payments concept of residence of both the reporting bank and its counterparty could not be applied. In this paper we analyze only cross-country relationships, so we exclude these groupings from the database as well as a position of international financial organizations.

As a result, we obtained a database covering 22 countries that have bank foreign claims and 198 countries that have obligations for the end of 1Q 2015. Thus, the network considered on the basis of the data includes all information about the international borrowings, except for transactions between countries which do not report. According to BIS Statistical Bulletin, our network covers about 94% of total foreign claims and other potential exposures on an ultimate risk basis.

The important aspect of the analysis is the choice of the critical loan amount threshold level for each country. One possible way to define it is to follow the recommendations of the Basel Committee (BCBS, 2013) on large exposure limits (25% of the Tier 1 capital). At the international level, when we deal with banking system's borrowings, choosing an appropriate threshold level (critical loan amount) is not so obvious exercise. We decided to put on the edges of the network not loans, but the value measured by the ratio of loans to the gross domestic product (GDP) of the lending country in order to take into account the relative size of the loan. In this paper the nominal GDP is used. However, GDP measure can be replaced by the total banking system assets or capital estimations. So, suppose that the threshold $q$ be 10% of the nominal GDP.

[g] According to Eurostat government finance statistics (http://ec.europa.eu/eurostat/statistics-explained/index.php/Government_expenditure_on_general_public_services)general government includes all institutional units whose output is intended for individual and collective consumption and mainly financed by compulsory payments made by units belonging to other sectors, and/or all institutional units principally engaged in the redistribution of national income and wealth. The general government sector is subdivided into four subsectors: central government, state government, local government, and social security funds.

Another important issue here is to assign grades to the direct influence values for the threshold procedure. In Table 20 we propose the following system of grades which in our opinion is reasonable for this case.

**Table 20.** Grades for direct influence values

| Grade | Condition | Description |
|---|---|---|
| 7 | $c_{ij} = 1$ | ultimately high influence |
| 6 | $0.92 \leq c_{ij} < 1$ | very high influence (explanation similar to the capital adequacy ratio for banks, when the loss of more than 0.92% of assets will lead to the bank's capital fall below zero) |
| 5 | $0.85 \leq c_{ij} < 0.92$ | high influence (according to the upper value of the capital adequacy ratio standard procedure of Bank of Russia) |
| 4 | $0.75 \leq c_{ij} < 0.85$ | average influence |
| 3 | $0.5 \leq c_{ij} < 0.75$ | moderate influence |
| 2 | $0.25 \leq c_{ij} < 0.5$ | low influence |
| 1 | $0 < c_{ij} < 0.25$ | very low influence |
| 0 | $c_{ij} = 0$ | no influence |

The highest grade corresponds to the highest influence value, the lowest grade indicates the absence of influence between elements.

Thus, we can calculate the influence value of each borrower according to our methodology. Table 21 contains a list of countries that were in the TOP-10 by one of LRIC indices. Graphical representation of the modeling network shown on Figure 7. We have also calculated the Key Borrower Index and compared the results with our methods.

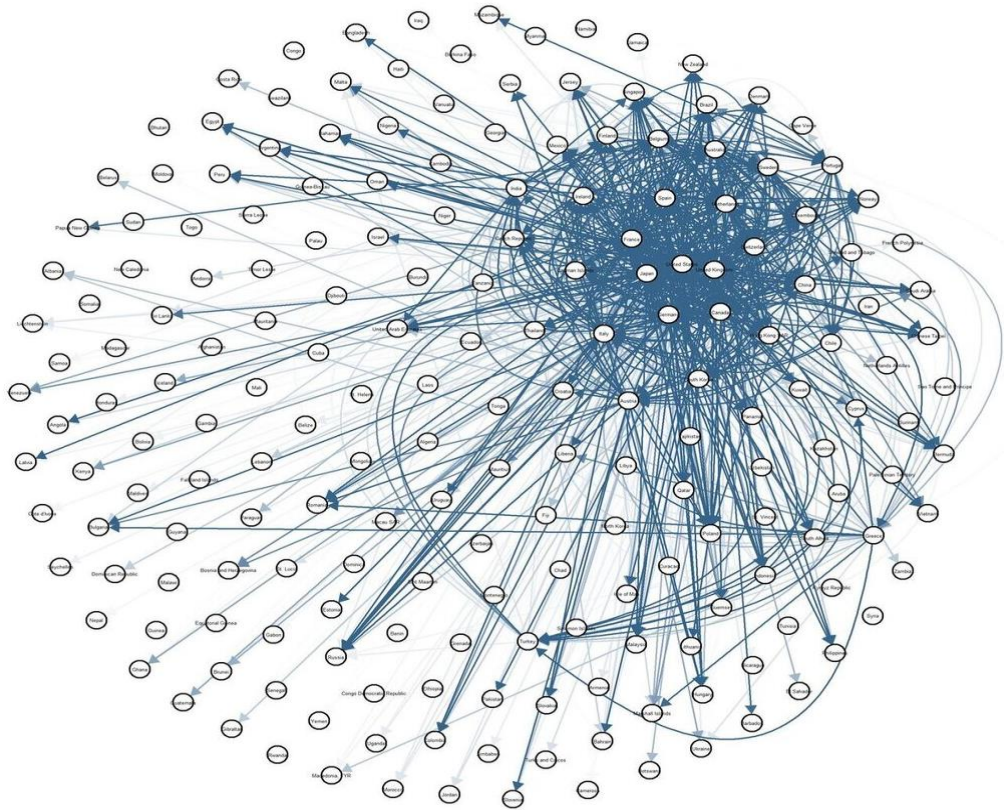

**Figure 7**. A graphical representation of banking foreign claims network

The countries with the largest value of the index are considered as the most pivotal/interconnected ones in the market. All five versions of LRIC (SumPaths, MaxPath, MaxMin, Simul, MaxT, MultT) give us almost similar rankings, whereas LRIC based on simulations demonstrates some differences. However, the main differences start from the middle of the TOP-10 countries. TOP-2 positions are stable according to all methods and occupied by the United States of America (USA) and Hong Kong. LRIC index based on simulations also groups in the ranking some countries forming regional clusters (Scandinavian, Baltic countries, Australia and New Zeeland).

The results allowed us to obtain two types of countries.

First of all, the highest ratings are typical for large and strong economies such as USA, UK and China. They have developed financial systems with high level of trustworthiness and sovereign ratings. As a result, their financial products (banking deposits or securities) attract a large number of investors. These results are in the line with findings of (IMF, 2015) and could be a good basis for "too big to fail" policy, when financial sectors of these countries could be a source of global systemic risk and should be more closely monitored.

**Table 21.** Power of countries as borrowers

| Country | KBI | LRIC indices | | | | | | Rank | | | | | | |
| | | Sum Paths | Max Path | MaxMin | Simul | MaxT | MultT | KBI | Sum Paths | Max Path | MaxMin | Simul | MaxT | MultT |
|---|---|---|---|---|---|---|---|---|---|---|---|---|---|---|
| United States | 0.542 | 0.083 | 0.133 | 0.080 | 0.540 | 0.1493 | 0.1115 | 1 | 1 | 1 | 1 | 1 | 1 | 1 |
| Hong Kong SAR | 0.051 | 0.073 | 0.078 | 0.052 | 0.149 | 0.0924 | 0.0651 | 4 | 2 | 2 | 2 | 2 | 2 | 2 |
| China | 0.004 | 0.061 | 0.053 | 0.044 | 0.004 | 0.0781 | 0.0443 | 19 | 3 | 4 | 3 | 16 | 4 | 4 |
| United Kingdom | 0.090 | 0.060 | 0.059 | 0.040 | 0.064 | 0.0694 | 0.0477 | 2 | 4 | 3 | 4 | 3 | 3 | 3 |
| Singapore | 0.002 | 0.040 | 0.029 | 0.028 | 0.004 | 0.0476 | 0.0237 | 25 | 5 | 8 | 6 | 17 | 7 | 7 |
| Cayman Islands | 0.005 | 0.039 | 0.042 | 0.031 | 0.008 | 0.0521 | 0.0326 | 17 | 6 | 5 | 5 | 14 | 5 | 5 |
| Brazil | 0.006 | 0.039 | 0.031 | 0.025 | 0.016 | 0.0310 | 0.0187 | 15 | 7 | 7 | 7 | 9 | 10 | 10 |
| Luxembourg | 0.010 | 0.035 | 0.026 | 0.019 | 0.010 | 0.0304 | 0.0214 | 11 | 8 | 9 | 12 | 13 | 8 | 8 |
| Poland | 0.037 | 0.033 | 0.018 | 0.017 | 0.003 | 0.0285 | 0.0135 | 6 | 9 | 17 | 16 | 18 | 14 | 14 |
| Germany | 0.037 | 0.029 | 0.033 | 0.022 | 0.010 | 0.0382 | 0.0271 | 5 | 10 | 6 | 9 | 12 | 6 | 6 |
| Mexico | 0.012 | 0.0283 | 0.0251 | 0.0193 | 0.041 | 0.0303 | 0.0178 | 10 | 11 | 10 | 10 | 5 | 11 | 11 |
| Czech Republic | 0.029 | 0.0233 | 0.0202 | 0.0191 | 0.0136 | 0.0277 | 0.0160 | 8 | 13 | 14 | 14 | 10 | 13 | 13 |
| Japan | 0.059 | 0.0180 | 0.0237 | 0.0191 | 0.043 | 0.0328 | 0.0196 | 3 | 17 | 11 | 11 | 4 | 9 | 9 |
| Norway | 0.005 | 0.0103 | 0.0091 | 0.0075 | 0.017 | 0.0107 | 0.0076 | 16 | 28 | 31 | 31 | 8 | 27 | 27 |
| Finland | 0.005 | 0.009 | 0.009 | 0.009 | 0.017 | 0.0105 | 0.0073 | 18 | 30 | 32 | 32 | 7 | 29 | 29 |
| Denmark | 0.009 | 0.009 | 0.009 | 0.008 | 0.017 | 0.0099 | 0.0071 | 13 | 31 | 33 | 33 | 6 | 30 | 30 |
| Italy | 0.034 | 0.0166 | 0.0208 | 0.0175 | 0.0124 | 0.0261 | 0.0169 | 7 | 18 | 13 | 13 | 11 | 12 | 12 |
| Austria | 0.024 | 0.012 | 0.014 | 0.015 | 0.000 | 0.0194 | 0.0106 | 9 | 26 | 22 | 22 | 54 | 20 | 20 |

However, in contrast to the previous works (Aleskerov et al. 2014), we can identify a group of countries that are not so large in terms of size of the economy, but also received the highest LRIC values. Countries like Hong Kong, Cayman Islands, Singapore and Luxembourg could be good examples of "too interconnected to fail" economies. Due to their attractive business environment, well-developed infrastructure, human capital and positive reputation, these countries stimulate investors place their assets in their financial systems, which makes these countries important borrowers. The appearance of these countries in the top ranking does not look normal at first sight, but it is in line with our initial hypothesis that the greatest influence must have not only the largest market participants, but also the most interconnected ones. In other words, for these countries each individual cash flow is not so significant, but their combination can be critical for the stability of the financial system as a whole. For example, in the case of the elimination of a country from the network, we will most likely not see a chain of cascading failures (because of volumes of interaction to each country are not so great), but it will lead to redirecting financial flows on the other countries that will affect the overall financial stability.

In this regard, there is an interesting question about the sensitivity of the results to changes in the network structure. According to our estimates, LRIC method allows determining the key elements in networks of any configuration, and can also be used to analyze the dynamics of the network configuration as well.

We have also estimated the level of country interconnectedness using a broad range of existing centrality measures: weighted degree centrality, closeness centrality, betweenness centrality, PageRank and eigenvector centrality. These measures are described in (von Peter, 2007), (Barrat et al., 2004) and we follow a very similar logic. More detailed description of the methodology is presented in the Section 2.

The results of the centrality indices calculations are shown in Table 22.

In order to compare rankings, we used a correlation analysis. Since the position in the ranking is a rank variable, to assess the consistency of different orderings other than traditional Pearson coefficient rank correlation coefficients should be used. In our work it is applied the idea of Kendall metrics (Kendall, 1970), that counts the number of pairwise disagreements between two ranking lists. Also we used Goodman and Kruskal $\gamma$ rank coefficient, which shows the similarity of the orderings of the data when ranked by each of the quantities (Goodman, Kruskal, 1954). This coefficient is calculated as $\gamma = \frac{N_S - N_D}{N_S + N_D}$, where $N_S$ is the number of pairs of cases ranked in the same order on both variables (number of concordant pairs) and $N_D$ is the number of pairs of cases ranked in reversed order on both variables (number of reversed pairs).

The results are provided below (Tables 22-23).

Table 22. Rankings by centrality measures

| Name | WInDeg | WOutDeg | WDeg | WDDif | Betw | Clos | PageRank | EigenVec |
|---|---|---|---|---|---|---|---|---|
| UnitedStates | 1 | 3 | 1 | 198 | 2 | 1 | 1 | 1 |
| UnitedKingdom | 2 | 1 | 2 | 3 | 1 | 2 | 2 | 2 |
| Germany | 3 | 5 | 5 | 6 | 14 | 58 | 3 | 5 |
| France | 4 | 4 | 1 | 2 | 16 | 105 | 4 | 4 |
| Japan | 6 | 2 | 2 | 1 | 11 | 54 | 5 | 3 |
| Netherlands | 10 | 9 | 5 | 9 | 15 | 151 | 6 | 9 |
| CaymanIslands | 5 | 59 | 4 | 197 | 22 | 157 | 7 | 10 |
| China | 9 | 61 | 3 | 195 | 24 | 133 | 8 | 14 |
| HongKong SAR | 8 | 97 | 9 | 196 | 25 | 125 | 9 | 13 |
| Italy | 7 | 12 | 13 | 13 | 8 | 79 | 10 | 11 |
| Spain | 11 | 6 | 15 | 4 | 4 | 158 | 11 | 8 |
| Canada | 16 | 8 | 14 | 7 | 10 | 61 | 12 | 6 |
| Luxembourg | 13 | 13 | 10 | 194 | 27 | 48 | 13 | 17 |
| Singapore | 14 | 14 | 6 | 192 | 28 | 64 | 14 | 16 |
| Brazil | 15 | 15 | 8 | 193 | 29 | 197 | 15 | 19 |
| Australia | 12 | 11 | 18 | 10 | 3 | 9 | 16 | 12 |
| Switzerland | 17 | 7 | 22 | 5 | 17 | 190 | 17 | 7 |
| Poland | 19 | 19 | 20 | 188 | 30 | 42 | 19 | 28 |
| Mexico | 21 | 21 | 11 | 191 | 33 | 116 | 21 | 24 |
| CzechRepublic | 22 | 22 | 7 | 180 | 34 | 70 | 22 | 41 |
| Belgium | 24 | 17 | 27 | 174 | 5 | 12 | 24 | 21 |
| India | 25 | 27 | 23 | 186 | 6 | 62 | 25 | 22 |
| Austria | 27 | 16 | 33 | 11 | 9 | 11 | 27 | 25 |
| Denmark | 28 | 28 | 17 | 189 | 35 | 108 | 28 | 34 |
| Sweden | 32 | 10 | 24 | 8 | 7 | 182 | 29 | 15 |
| Norway | 29 | 144 | 16 | 185 | 36 | 188 | 30 | 37 |
| Finland | 30 | 29 | 26 | 183 | 19 | 3 | 33 | 35 |

We can see that according to our estimations, the rankings of LRIC indices are highly related to the results of the PageRank. This fact is confirmed by both our correlation coefficients (Kendall $\tau$ and Goodman, Kruskal $\gamma$-coefficient). It should be noted that the weighted degree centrality also gives us a similar rankings. As for other centrality measures, their correlation coefficients are relatively high except the betweenness centrality and weighted out-degree centrality measures for which the correlation coefficients is less than 0,5 (Kendall $\tau$) or less than 0,4 ($\gamma$-coefficient).

However, although the correlation coefficients of most centralities measures and LRIC indices are quite high, it was shown in Table 22 that in contrast to LRIC indices classical centrality measures are worse in detecting systemically important countries of the second group (e.g. Cayman Island, Luxembourg, Hong Kong).

**Table 23.** Kendall $\tau$-coefficient

| | WInDeg | WOutDeg | WDeg | WDDif | Clos | PageRank | EigenVec | Betw | Simul | SumPaths | MaxPath | MaxMin | MaxT | MultT |
|---|---|---|---|---|---|---|---|---|---|---|---|---|---|---|
| **WInDeg** | - | 0.434 | 0.963 | -0.664 | 0.978 | 0.984 | 0.963 | 0.546 | 0.735 | 0.824 | 0.836 | 0.864 | 0.874 | 0.862 |
| **WOutDeg** | | - | 0.548 | -0.354 | 0.863 | 0.965 | 0.976 | 0.455 | 0.314 | 0.812 | 0.796 | 0.814 | 0.822 | 0.804 |
| **WDeg** | | | - | -0.781 | 0.908 | 0.987 | 0.974 | 0.464 | 0.911 | 0.926 | 0.93 | 0.918 | 0.936 | 0.956 |
| **WDDif** | | | | - | -0.564 | -0.806 | -0.794 | -0.386 | -0.654 | -0.698 | -0.674 | -0.688 | -0.656 | -0.647 |
| **Clos** | | | | | - | 0.898 | 0.922 | 0.483 | 0.845 | 0.869 | 0.87 | 0.843 | 0.874 | 0.865 |
| **PageRank** | | | | | | - | 0.951 | 0.438 | 0.913 | 0.937 | 0.939 | 0.933 | 0.952 | 0.936 |
| **EigenVec** | | | | | | | - | 0.456 | 0.869 | 0.897 | 0.897 | 0.866 | 0.963 | 0.942 |
| **Betw** | | | | | | | | - | 0.355 | 0.344 | 0.359 | 0.353 | 0.341 | 0.325 |
| **Simul** | | | | | | | | | - | 0.966 | 0.967 | 0.965 | 0.964 | 0.955 |
| **SumPaths** | | | | | | | | | | - | 0.998 | 0.987 | 0.96 | 0.943 |
| **MaxPath** | | | | | | | | | | | - | 0.988 | 0.985 | 0.976 |
| **MaxMin** | | | | | | | | | | | | - | 0.966 | 0.959 |
| **MaxT** | | | | | | | | | | | | | - | 1.000 |
| **MultT** | | | | | | | | | | | | | | - |

**Table 24.** Goodman, Kruskal $\gamma$-coefficient

| | WInDeg | WOutDeg | WDeg | WDDif | Clos | PageRank | EigenVec | Betw | Simul | SumPaths | MaxPath | MaxMin | MaxT | MultT |
|---|---|---|---|---|---|---|---|---|---|---|---|---|---|---|
| **WInDeg** | - | 0.384 | 0.754 | -0.487 | 0.796 | 0.753 | 0.721 | 0.396 | 0.587 | 0.69 | 0.674 | 0.687 | 0.699 | 0.659 |
| **WOutDeg** | | - | 0.469 | -0.304 | 0.639 | 0.723 | 0.705 | 0.388 | 0.363 | 0.735 | 0.758 | 0.801 | 0.814 | 0.784 |
| **WDeg** | | | - | -0.654 | 0.745 | 0.903 | 0.88 | 0.378 | 0.73 | 0.767 | 0.772 | 0.743 | 0.786 | 0.76 |
| **WDDif** | | | | - | -0.456 | -0.789 | -0.628 | -0.265 | -0.547 | -0.523 | -0.564 | -0.599 | -0.503 | -0.526 |
| **Clos** | | | | | - | 0.728 | 0.763 | 0.397 | 0.654 | 0.684 | 0.684 | 0.657 | 0.754 | 0.798 |
| **PageRank** | | | | | | - | 0.828 | 0.354 | 0.739 | 0.791 | 0.789 | 0.768 | 0.789 | 0.789 |
| **EigenVec** | | | | | | | - | 0.371 | 0.673 | 0.719 | 0.719 | 0.681 | 0.695 | 0.684 |
| **Betw** | | | | | | | | - | 0.289 | 0.278 | 0.291 | 0.285 | 0.256 | 0.289 |
| **Simul** | | | | | | | | | - | 0.887 | 0.887 | 0.874 | 0.864 | 0.876 |
| **SumPaths** | | | | | | | | | | - | 0.976 | 0.924 | 0.915 | 0.919 |
| **MaxPath** | | | | | | | | | | | - | 0.93 | 0.925 | 0.936 |
| **MaxMin** | | | | | | | | | | | | - | 0.935 | 0.928 |
| **MaxT** | | | | | | | | | | | | | - | 1.000 |
| **MultT** | | | | | | | | | | | | | | - |

# Conclusion

Network approach can be applied to different segments of the financial system in order to characterize a systemic risk. In our work network models of systemic risk have been applied to a specific section of the financial market – international credit market.

We explore two approaches to measure systemic importance: measurement of long-range interactions' intensities based on paths, and measurement of long-range interactions' intensities based on simulations. This network model aims to help regulators identify the financial elements that are too big or too interconnected to fail during any specific crisis. The proposed methodology allows to identify countries, which at first sight do not have high level of systemic importance, but have a significant impact on the stability of the system as a whole. Also LRIC index based on simulations could be a useful instrument for identification of regional financial clusters.

We carried out estimations for hypothetical examples and presented an empirical analysis of cross-border country exposures to demonstrate the feasibility of the proposed methodology. The empirical results based on our methodology are in line with the conclusions made by IMF and other international financial institutions. In addition, these results draw our attention to the importance of countries, which due to their intermediation role in the global finances can have a strong influence on the stability of the entire system.

# Appendix 1. Calculation of centrality measures, Key Borrower Index and long-range interaction indices for the Numerical Example 1.

## 1.1 Calculation of centrality measures

The calculation of classical centrality measures was performed in R 3.2.2 software package with the use of embedded functions (see Table 25).

**Table 25.** List of functions for centrality measures (g is an input graph)

| Centrality measure | Function |
|---|---|
| Weighted In-degree | strength(g,mode="in") |
| Weighted Out-degree | strength(g,mode="out") |
| Weighted Degree Difference | strength(g,mode="out")-strength(g,mode="in") |
| Weighted Degree | strength(g,mode="total") |
| Closeness, in | closeness(g,mode="in") |
| Closeness, out | closeness(g,mode="out") |
| Betweenness | betweenness(g) |
| Eigenvector | evcent(g, directed=TRUE) |
| PageRank | page.rank(g, directed=TRUE) |

## 1.2 Calculation of Key Borrower Index

According to (Aleskerov et al., 2014) we consider each lender separately, calculate the influence of each borrower to the particular lender and aggregate the results over all lenders with respect to the size of the loans. For example, the lender 1 gave $1000 ($500+$100+$400), total sum of money which was borrowed in the system is equal to $3660 ($1000+$200+ +$150+$60+$1100+$850+$150), so the lender 1 weight is equal to 0.27

**Table 26.** Key Borrower Index for Numerical Example 1

| Lenders | Key Borrower Index | | | | | | | | | | Weight |
|---|---|---|---|---|---|---|---|---|---|---|---|
| | 1 | 2 | 3 | 4 | 5 | 6 | 7 | 8 | 9 | 10 | |
| L=1 | 0 | 0.556 | 0 | 0 | 0.444 | 0 | 0 | 0 | 0 | 0 | 0.27 |
| L=2 | 0 | 0 | 0 | 0 | 0 | 0.654 | 0 | 0 | 0.346 | 0 | 0.05 |
| L=3 | 0 | 0 | 0 | 0 | 0 | 1 | 0 | 0 | 0 | 0 | 0.04 |
| L=4 | 0 | 0 | 0 | 0 | 0 | 1 | 0 | 0 | 0 | 0 | 0.02 |
| L=5 | 0 | 0 | 0 | 0 | 0 | 0.875 | 0.063 | 0.063 | 0 | 0 | 0.30 |
| L=7 | 0 | 0 | 0 | 0 | 0 | 0 | 0 | 0 | 0.706 | 0.294 | 0.27 |
| L=8 | 0 | 0 | 0 | 0 | 0 | 0 | 0 | 0 | 0 | 1 | 0.04 |
| **Total** | **0** | **0.152** | **0** | **0** | **0.121** | **0.356** | **0.019** | **0.019** | **0.212** | **0.121** | |

### 1.3 Calculation of Long-Range Interaction Centrality indices

For Numerical Example 1 we consider all possible groups of borrowers of size less or equal to 5 while the parameter *s* that defines how many "layers" are examined for each lender is undefined.

*Long-Range Interaction Centrality indices based on paths*

Firstly, let us construct a matrix $C = [c_{ij}]$ for threshold value *q*=25% according to Section 3.1. The corresponding network is shown on Figure 8.

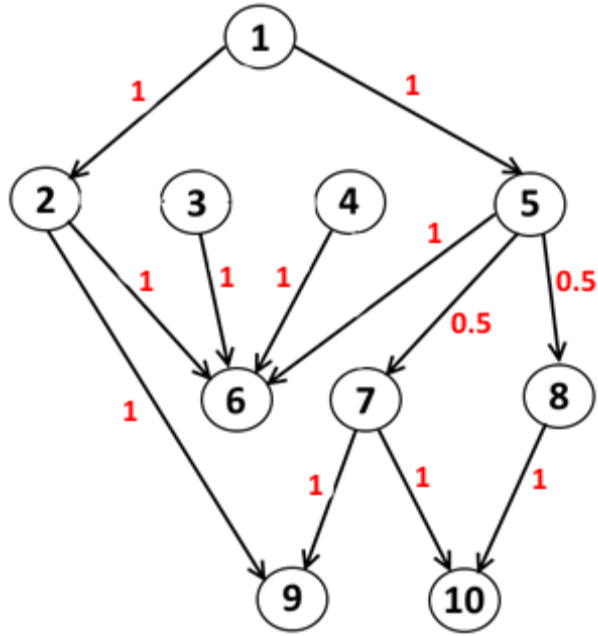

**Figure 8**. The network for matrix $C$

The matrix C is used to evaluate the long-range influence between elements of the system. To do it, several methods of indirect influence evaluation was proposed. Thus, for each method we can construct the matrix $C^*$ which represents the influence and is used for the aggregation of influences into a single vector with respect to the weights. For the Numerical Example 1 the matrix $C^*$ is the same for all methods. The results are provided in Table 27.

**Table 27.** Matrix $C^*$ for the Numerical Example 1 (paths)

|  | **1** | **2** | **3** | **4** | **5** | **6** | **7** | **8** | **9** | **10** | **Weights** |
|---|---|---|---|---|---|---|---|---|---|---|---|
| **1** | 0 | 1 | 0 | 0 | 1 | 1 | 0.5 | 0.5 | 1 | 0.5 | 0.27 |
| **2** | 0 | 0 | 0 | 0 | 0 | 1 | 0 | 0 | 1 | 0 | 0.05 |
| **3** | 0 | 0 | 0 | 0 | 0 | 1 | 0 | 0 | 0 | 0 | 0.04 |
| **4** | 0 | 0 | 0 | 0 | 0 | 1 | 0 | 0 | 0 | 0 | 0.02 |
| **5** | 0 | 0 | 0 | 0 | 0 | 1 | 0.5 | 0.5 | 0.5 | 0.5 | 0.30 |
| **6** | 0 | 0 | 0 | 0 | 0 | 0 | 0 | 0 | 0 | 0 | 0.00 |
| **7** | 0 | 0 | 0 | 0 | 0 | 0 | 0 | 0 | 1 | 1 | 0.27 |
| **8** | 0 | 0 | 0 | 0 | 0 | 0 | 0 | 0 | 0 | 1 | 0.04 |
| **9** | 0 | 0 | 0 | 0 | 0 | 0 | 0 | 0 | 0 | 0 | 0.00 |
| **10** | 0 | 0 | 0 | 0 | 0 | 0 | 0 | 0 | 0 | 0 | 0.00 |
| **Total** | 0.00 | 0.27 | 0.00 | 0.00 | 0.27 | 0.69 | 0.29 | 0.29 | 0.75 | 0.60 | |
| **Total (normalized)** | 0 | 0.09 | 0 | 0 | 0.09 | **0.22** | 0.09 | 0.09 | **0.23** | **0.19** | |

*Long-Range Interaction Centrality indices based on simulations*

Firstly, let us construct a matrix $C = [c_{ij}]$ for threshold value $q$=25% according to the Section 3.2. A graphical representation of the matrix is shown on Figure 9.

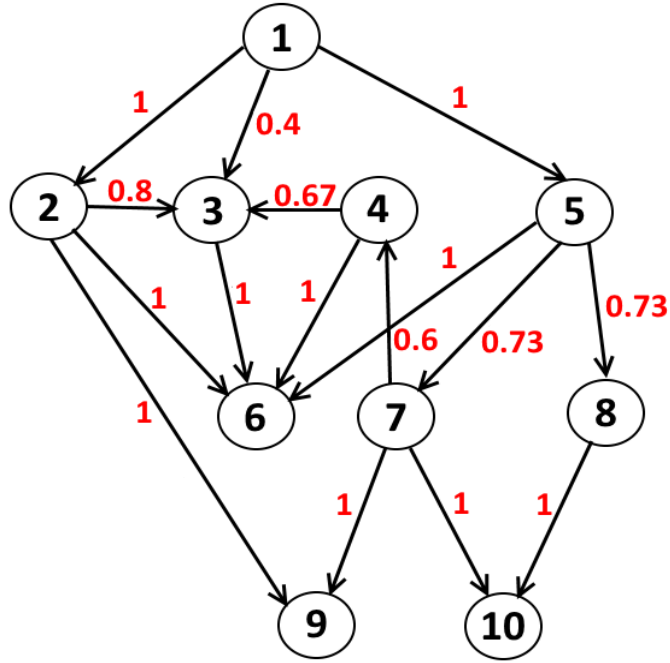

**Figure 9.** The network for matrix $C$

The matrix C is used to evaluate the long-range influence between elements of the system based on the idea of simulations. To do it, 5000 possible combinations of borrowers were considered to construct a new matrix $C^*$. Thus, for each method the matrix $C^*(s) = [c_{ij}^*(s)]$, which shows in what percentage of cases the borrower $i$ could not return its loans if we assume that the borrower $j$ is not capable to return his own loans, was constructed. The results can be aggregated into a single vector with respect to the weights (see Table 28).

**Table 28.** Matrix $C^*$ for the Numerical Example 1 (simulations)

| | 1 | 2 | 3 | 4 | 5 | 6 | 7 | 8 | 9 | 10 | Weights |
|---|---|---|---|---|---|---|---|---|---|---|---|
| **1** | 0 | 1 | 0 | 0 | 1 | 1 | 0.4 | 0.2 | 1 | 1 | 0.27 |
| **2** | 0 | 0 | 0 | 0 | 0 | 1 | 0 | 0 | 1 | 0 | 0.05 |
| **3** | 0 | 0 | 0 | 0 | 0 | 1 | 0 | 0 | 0 | 0 | 0.04 |
| **4** | 0 | 0 | 0 | 0 | 0 | 1 | 0 | 0 | 0 | 0 | 0.02 |
| **5** | 0 | 0 | 0 | 0 | 0 | 1 | 0.4 | 0.58 | 0.47 | 1 | 0.30 |
| **6** | 0 | 0 | 0 | 0 | 0 | 0 | 0 | 0 | 0 | 0 | 0.00 |
| **7** | 0 | 0 | 0 | 0 | 0 | 0 | 0 | 0 | 1 | 1 | 0.27 |
| **8** | 0 | 0 | 0 | 0 | 0 | 0 | 0 | 0 | 0 | 1 | 0.04 |
| **9** | 0 | 0 | 0 | 0 | 0 | 0 | 0 | 0 | 0 | 0 | 0.00 |
| **10** | 0 | 0 | 0 | 0 | 0 | 0 | 0 | 0 | 0 | 0 | 0.00 |
| **Total** | 0 | 0.27 | 0 | 0 | 0.27 | 0.69 | 0.23 | 0.23 | 0.74 | 0.89 | |
| **Total (normalized)** | 0 | 0.085 | 0 | 0 | 0.085 | **0.211** | 0.072 | 0.071 | **0.216** | **0.261** | |

# Appendix 2. Calculation of centrality measures and Key Borrower Index for Numerical Example 2.

## 2.1 Calculation of centrality measures

The calculation of classical centrality measures for the Numerical Example 2 was performed in R 3.2.2 software package as it was done for Numerical Example 1 (see Appendix 1).

## 2.2 Calculation of Key Borrower Index

According to (Aleskerov et al., 2014) we consider each lender separately, calculate the influence of each borrower to the particular lender and aggregate the results over all lenders with respect to the size of the loans. For example, the lender 1 gave $100 ($60+$16+$24), total sum of money which was borrowed in the system is equal to $900 ($100+$100+$100+$100+ +$100+$100+$100+$100+$100), so the lender 1 weight is equal to 0.11

**Table 29.** Key Borrower Index for Numerical Example 2

| Lenders | Key Borrower Index | | | | | | | | | | | Weight |
|---------|---|---|---|---|---|---|---|---|---|---|---|--------|
|         | 1 | 2 | 3 | 4 | 5 | 6 | 7 | 8 | 9 | 10 | 11 | |
| L=1  | 0 | 0.82 | 0.05 | 0.13 | 0 | 0 | 0 | 0 | 0 | 0 | 0 | 0.11 |
| L=2  | 0 | 0 | 0 | 0 | 0.02 | 0.90 | 0 | 0.08 | 0 | 0 | 0 | 0.11 |
| L=3  | 0 | 0.08 | 0 | 0.84 | 0.08 | 0 | 0 | 0 | 0 | 0 | 0 | 0.11 |
| L=4  | 0 | 0 | 0 | 0 | 0.03 | 0 | 0.89 | 0 | 0.08 | 0 | 0 | 0.11 |
| L=6  | 0 | 0 | 0 | 0 | 0 | 0 | 0 | 0 | 0 | 0 | 1 | 0.11 |
| L=7  | 0 | 0 | 0 | 0 | 0 | 0 | 0 | 0 | 0 | 0 | 1 | 0.11 |
| L=8  | 0 | 0 | 0 | 0 | 0 | 0 | 0 | 0 | 0 | 0 | 1 | 0.11 |
| L=9  | 0 | 0 | 0 | 0 | 0 | 0 | 0 | 0 | 0 | 0 | 1 | 0.11 |
| L=10 | 1 | 0 | 0 | 0 | 0 | 0 | 0 | 0 | 0 | 0 | 0 | 0.11 |
| **Total** | **0.11** | **0.10** | **0.01** | **0.11** | **0.01** | **0.10** | **0.10** | **0.01** | **0.01** | **0.00** | **0.44** | |

**Fuad Aleskerov**
National Research University Higher School of Economics (HSE), International Laboratory of Decision Choice and Analysis, V.A. Trapeznikov Institute of Control Sciences of Russian Academy of Sciences (ICS RAS), Moscow, alesk@hse.ru

**Any opinions or claims contained in this Working Paper do not necessarily reflect the views of HSE.**